\documentclass[useAMS,usenatbib]{mn2e}
\usepackage{deluxetable}
\usepackage{epsf}
\usepackage{amsmath}
\usepackage{subfigure}
\usepackage{url}
\usepackage{color}

\newcommand{\smyr} {\msun \mbox{ yr}^{-1}}

\newcommand{\rvir}{R_{\rm vir}}

\newcommand{\hkpc}{h^{-1}{\rm kpc}}
\newcommand{\hmpc}{h^{-1}{\rm Mpc}}

\newcommand{\kms}{\;{\rm km}\,{\rm s}^{-1}}

\newcommand{\msun}{M_{\odot}}

\newcommand{\rhodotstar}{\dot{\rho}_*}

\newcommand{\rhogas}{\rho_g}

\newcommand{\dlls}{\Delta_{\rm{lls}}}

\newcommand{\aap}{A\&A}
\newcommand{\apjs}{ApJS}
\newcommand{\apj}{ApJ}
\newcommand{\apjl}{ApJL}
\newcommand{\aj}{AJ}

\newcommand{\mnras}{MNRAS}
\newcommand{\prd}{PhRvD}

\newcommand{\nat}{Nature}

\newcommand{\fesc}{f_{\mathrm{esc}}}
\newcommand{\fescfive}{f_{\mathrm{esc,5}}}

\newcommand{\tes}{\tau_{\mathrm{es}}}
\newcommand{\ta}{\tau_\alpha}

\newcommand{\NOI}{N_{\mathrm{OI}}}
\newcommand{\NHI}{N_{\mathrm{HI}}}

\begin{document}

\title[OI-Absorbing Halos]{The Host Halos of OI Absorbers in the Reionization Epoch}

\author[K.\ Finlator et al.]{
\parbox[t]{\textwidth}{\vspace{-1cm}
Kristian Finlator$^{1,10}$, Joseph A.\ Mu{\~n}oz$^2$, B.\ D.\ Oppenheimer$^{3,4}$, S.\ Peng Oh$^5$, Feryal \"{O}zel$^6$, Romeel Dav\'e$^{6,7,8,9}$}
\\\\$^1$ Dark Institute of Cosmology, Niels Bohr Institute, University of Copenhagen, Copenhagen, Denmark
\\$^2$ University of California Los Angeles, Departmen of Physics and Astronomy, CA 90095, USA
\\$^3$ Leiden Observatory, Leiden University, PO Box 9513, Leiden, Netherlands
\\$^4$ CASA, Department of Astrophysical and Planetary Sciences, University of Colorado, 389-UCB, Boulder, CO 80309, USA
\\$^5$ Department of Physics, University of California Santa Barbara, Santa Barbara, CA 93106, USA
\\$^6$ Astronomy Department, University of Arizona, Tucson, AZ 85721, USA
\\$^7$ University of the Western Cape, Bellville, Cape Town 7535, South Africa
\\$^8$ South African Astronomical Observatories, Observatory, Cape Town 7525, South Africa
\\$^9$ African Institute for Mathematical Sciences, Muizenberg, Cape Town 7545, South Africa
\\$^{10}$ kfinlator@dark-cosmology.dk
\author[OI-absorbing Halos]{K.\ Finlator, J.\ A.\ Mu{\~n}oz, B.\ D.\ Oppenheimer, S.\ Peng Oh, F.\ \"{O}zel, \& R.\ Dav\'e}
}

\maketitle

\begin{abstract}
We use a radiation hydrodynamic simulation of the hydrogen reionization epoch 
to study OI absorbers at $z\sim6$.  The intergalactic medium (IGM) is reionized 
before it is enriched, hence OI absorption originates within dark matter halos.  
The predicted abundance of OI absorbers is in reasonable agreement with 
observations.  At $z=10$, $\approx70\%$ of sightlines through atomically-cooled 
halos encounter a visible ($\NOI > 10^{14}$cm$^{-2}$) column.  
Reionization ionizes and removes gas from halos less massive than 
$10^{8.4}\msun$, but 20\% of sightlines through more massive halos encounter 
visible columns even at $z=5$.  The mass scale of absorber host halos is 
10--100$\times$ smaller than the halos of Lyman break galaxies and 
Lyman-$\alpha$ emitters, hence absorption probes the dominant ionizing 
sources more directly.  OI absorbers have neutral hydrogen columns of 
$10^{19}$--$10^{21}$cm$^{-2}$, suggesting a close resemblance between 
objects selected in OI and HI absorption.  Finally, the absorption in 
the foreground of the $z=7.085$ quasar 
ULASJ1120+0641 cannot originate in a dark matter halo because halo gas at 
the observed HI column density is enriched enough to violate the upper
limits on the OI column.  By contrast, gas at less than one third the cosmic 
mean density satisfies the constraints.  Hence the foreground 
absorption likely originates in the IGM.
\end{abstract}

\begin{keywords}
cosmology: theory --- galaxies: halos --- galaxies: high-redshift --- galaxies: formation --- galaxies: evolution --- quasars: absorption lines
\end{keywords}

\section{Introduction} \label{sec:intro}
Mapping out the progress of hydrogen reionization and understanding the 
nature of the sources that drove it constitute two of the central 
challenges that Astronomy will confront over the coming 
decade~\citep{a2010}.  The cosmic microwave background constrains 
reionization to be roughly 50\% complete at some point between 
$z=$9--11.8, although the results depend on the shape of the 
assumed reionization history~\citep{hin12,mit12,pan11}.  The classic approach 
of measuring the neutral hydrogen fraction directly from the 
Lyman-$\alpha$ forest becomes increasingly difficult at redshifts beyond 
$z=6$ owing to the fact that Lyman-$\alpha$ absorption saturates for
neutral hydrogen fractions in excess of $10^{-3}$~\citep{fan02}.  In 
response to this challenge, a number of alternative techniques have 
been developed involving the abundance of Lyman-$\alpha$ 
emitters~\citep{ouc10,treu12} or Lyman break 
galaxies~\citep{mun11,fink12,rob13,oes13}, the statistics of dark pixels or gaps in the 
Lyman-$\alpha$ forest~\citep{mes10,mcg11}, or the presence of damping wings in quasar 
spectra~\citep{bol11,sch13}.  Each of these approaches combines unique strengths 
and weaknesses, hence it is necessary to consider a diverse variety of approaches 
together in order to overcome the weaknesses of any individual one.

One probe that has received relatively little attention involves the study
of low-ionization metal absorbers~\citep{oh02, fur03}.  If diffuse regions 
of the pre-reionization intergalactic medium (IGM) were enriched with
metals whose ionization potential is similar to that of hydrogen, then 
it may be possible to measure the ionization state of the metals directly 
and use this to trace the ionization state of the IGM as a whole.  
Recently,~\citet{bec11} searched for low-ionization metal absorbers
in moderate- and high-resolution spectra of 17 quasars at redshifts 
5.8--6.4.  They found that the abundance of systems at $z\sim6$ roughly 
matches the combined number density of 
damped Ly$\alpha$ systems (DLAs; $2\times10^{20} < \NHI/\mathrm{cm}^{-2}$)
and sub-DLAs ($10^{19} < \NHI/\mathrm{cm}^{-2} < 2\times10^{20}$) at 
$z\sim3$.  Furthermore, the velocity widths of the 
high-redshift absorbers are similar to those of the DLAs, although with
weaker equivalent widths.  The authors concluded that low-ionization metal 
absorbers trace low-mass halos rather than neutral regions in the 
diffuse IGM.

Modeling OI absorbers in order
to study the viability of this scenario requires a model that treats the 
inhomogeneous ionization and metal enrichment fields simultaneously.
In~\citet{opp09}, we used a 
cosmological hydrodynamic simulation that assumed a spatially-homogeneous 
extragalactic ultraviolet ionizing background (EUVB) to study metal 
absorbers in the reionization epoch.
The ionization field was adjusted in post-processing to consider scenarios 
in which there was no EUVB, a spatially-homogeneous 
EUVB, and an inhomogeneous model in which the EUVB at any 
point was dominated by the nearest galaxy.  It was found that the OI 
absorber abundance was dramatically overproduced in the absence of an 
EUVB, and underproduced under the assumption of an optically-thin 
EUVB or a simple model in which the EUVB at any point was 
governed by the nearest galaxy.  This work neglected two important
aspects of the radiation field: First, the clustered nature of 
ionizing sources means that the EUVB at any point is determined by 
the combined influence of many galaxies rather than just the nearest 
one~\citep{bar04,fur04a,fur04b,fur05}.  Second, dense sources acquire
a multiphase ionization structure consisting of an optically-thick
core and an optically-thin atmosphere.  Modeling the ionization front 
that separates these phases requires a spatial resolution of $~\sim1$ 
physical kpc~\citep{sch01,gne06,mcq11}, which was not achievable through the 
simple treatment adopted in~\citet{opp09}.  For these reasons, the 
spatial dependence of the assumed radiation field was incorrect.  
Hence while our previous study confirmed that there is enough oxygen 
to account for observations, the crude treatment of the EUVB 
meant that direct comparison with observations was 
preliminary.  

Here, we remedy these deficiencies by studying the nature of OI 
absorption using cosmological simulations in which the EUVB and the 
galaxies are modeled simultaneously and self-consistently.  We focus on OI 
absorbers because the abundance of oxygen leads to high OI columns while 
the proximity of its ionization potential to that of hydrogen means that
the neutral oxygen fraction can be obtained trivially from the neutral 
hydrogen fraction.  The goals of the current study are:
(1) To study the relative spatial distributions of enriched and ionized gas and
determine which portion of the IGM OI observations likely probe; (2) To
understand the impact of reionization on the sources of OI absorption; (3)
To compare the predicted and observed abundances of OI absorbers; and (4)
to compare the HI and OI absorption properties of halo gas in the 
reionization epoch.  Additionally, we will use our model to interpret 
observational constraints on the abundance of OI in the absorbing system 
that lies in the foreground of the $z=7.085$ quasar 
ULASJ1120+0641~\citep{mor11}.

In Section~\ref{sec:sims}, we introduce our simulations.  In Section~\ref{sec:ZxHI},
we explore the spatially-inhomogeneous ionization and chemical enrichment
fields in our simulations.  In Section~\ref{sec:obs}, we use insights from our
simulations to model the abundance of neutral oxygen absorbers as a function
of redshift and compare with observations.  We also compare the predicted HI
and OI absorption properties of reionization-epoch halos.  In 
Section~\ref{sec:discuss} we discuss our results with an eye toward future 
modeling efforts, and in Section~\ref{sec:sum} we summarize.

\section{Simulating Reionization and Enriched Outflows} \label{sec:sims}
\subsection{Simulations}\label{ssec:sims}
We use hydrodynamic simulations to model the inhomogeneous
ionization and metallicity fields.  These simulations are built on
the parallel N-body + smoothed particle hydrodynamics (SPH) code 
{\sc Gadget-2}~\citep{spr05} and include treatments for radiative 
cooling, star formation, and momentum-driven galactic outflows (except
for one simulation as we describe below).
We model the EUVB on-the-fly by solving the moments of the 
radiation transport equation on a Cartesian grid that is superposed on 
our simulation volume.  The ionizing emissivity within each cell is 
determined by the local star formation rate density, with a metallicity 
weighting based on the stellar population models of~\citet{sch03}.  The 
fraction of ionizing photons that escape into the IGM
varies depending on the simulation (see below).  The radiation and 
ionization fields are updated simultaneously using an iterative procedure. 
For details on all of these ingredients, see~\citet{fin11b,fin12}.

Three of the four simulations account for the ability for dense gas 
to acquire an optically-thick core on spatial scales beneath the 
resolution limit of our radiation transport solver.  We introduced 
this subgrid treatment in~\citet{fin12}, but we review it here as it
is a critical ingredient for modeling low-ionization metal absorbers. 

Directly resolving the ionization fronts that isolate optically-thick
regions requires a spatial resolution of $\sim1$ physical
kpc~\citep{sch01,gne06,mcq11}.  By contrast, our highest-resolution simulation
discretizes the radiation field using mesh cells that are 187.5$\hkpc$
wide (comoving).  While this allows us to model our volume's reionization 
history with $10^5$ cells, the resolution remains roughly a factor of 10 
too coarse to resolve Lyman limit systems (LLS; $\NHI>10^{17}$cm$^{-2}$).  
We overcome this limitation through 
a generalization of the~\citet{hae98} self-shielding scenario.  Each
SPH particle is exposed to an EUVB that is attenuated by an optical depth 
$\tau_\Gamma$ that varies with the local overdensity 
$\Delta\equiv\rho/\langle\rho\rangle$ as $\tau_\Gamma = (\Delta/\dlls)^b$.
The characteristic scale $\dlls$ is the overdensity of systems 
through which an optical depth of unity is expected under the assumption
that the gas is in hydrostatic equilibrium.  It depends on the local 
temperature, redshift, and the amplitude of the EUVB~\citep{sch01}, and
it grows from $\sim10$ at $z=10$ to $\sim100$ by $z=6$ (Figure
2 of~\citealt{fin12}).  We set the power-law slope $b=3$, although this 
choice does not affect the results significantly.  We also add the 
opacity of the self-shielded gas to the overall opacity field for self-consistency.  
Gas with $\Delta < \dlls$ sees an unattenuated EUVB.
This treatment yields an ionization field in which gas that is 
more than a few times more dense than $\dlls$ is neutral, in agreement 
with simulations that model the ionization field with higher resolution 
in a post-processing step~\citep{mcq11}.

Table~\ref{table:sims} shows our suite of simulations.  The naming
convention encodes the simulation parameters.  For example, the
r6n256wWwRT16d simulation subtends $6\hmpc$ (r6) using $2\times256^3$
particles (n256) with outflows (wW) and discretizes the radiation field
using $16^3$ cells (wRT16) including subgrid self-shielding (d).  For
all but the r6n256wWwRT simulation, the 
ionizing escape fraction varies with redshift as
\begin{eqnarray}\label{eqn:fesc}
\fesc = \left\{ 
\begin{array}{lc}
  \fescfive \left(\frac{1+z}{6}\right)^\kappa & z < 10 \\
  1.0 & z \geq 10
\end{array} \right.
\end{eqnarray}
Here, the normalization $\fescfive$ sets the escape fraction at $z=5$, which
we tune to match the observed ionizing emissivity at that 
redshift~\citep{kuh12a}.  The slope $\kappa$ controls how strongly $\fesc$ 
varies with redshift and is tuned to reach 1 at $z=10$.  These requirements
lead us to adopt $\fescfive=0.0519$ and $\kappa=4.8$ for the 
r6n256wWwRT16d and r9n384wWRT48d simulations.  The r6n256nWwRTd simulation 
is similar but does not include outflows.  Without outflows, the predicted 
star formation rate density is higher, hence we require a lower escape 
fraction in order to match observations; we adopt $\fescfive=0.0126$ and 
$\kappa=7.21$.  The r6n256wWwRT simulation does not include self-shielding 
and assumes a constant ionizing escape fraction $\fesc=0.5$.  
Note that our r9n384wWwRT48d run includes the same underlying
physics as the r6n256wWwRT16d run but 3.375 times more volume and a
finer radiation transport mesh, giving it the highest dynamic range that
we have modeled to date.  It required 71,000 CPU hours on 128 processors 
to reach $z=6$.  It is the fiducial simulation volume for the current study.

All simulations incorporate the same resolution such that the mass of a
halo with 100 dark matter and SPH particles is $1.4\times10^8\msun$, and the
gravitational softening length is $0.1$ kpc (Plummer equivalent; proper 
units at $z=6$).

We generate the initial density field using an~\citet{eis99} power 
spectrum at redshifts of 249 and 200 for simulations subtending 6 and 
9 $\hmpc$, respectively.  We initialize the IGM temperature and neutral 
hydrogen fraction to the values appropriate for each simulation's 
initial redshift as computed by {\sc recfast}~\citep{won08}, and 
we assume that helium is initially completely neutral.  All simulations 
assume a cosmology in which 
$\Omega_M=0.28$, $\Omega_\Lambda=0.72$, $\Omega_b = 0.046$, 
$h=0.7$, $\sigma_8 = 0.82$, and the index of the primordial power 
spectrum $n=0.96$.

The focus of our current work is the spatial distribution of neutral
oxygen.  Our simulations do not evolve the ionization state of oxygen 
on-the-fly because it contributes negligibly to the total opacity.
In order to compute the abundance of neutral oxygen, we combine in 
post-processing the predicted neutral 
hydrogen fraction and total oxygen abundance (which are both modeled 
on-the-fly) with the assumption that hydrogen and oxygen are in charge 
exchange equilibrium at the local gas temperature.  To do this, we 
use the expression~\citep{oh02}  
\begin{equation*}
\frac{\NOI}{N_{\mathrm{OII}}} = \frac{9}{8}\frac{\NHI}{N_{\mathrm{HII}}}\exp\left(\frac{\Delta E}{k_B T}\right),
\end{equation*}
where $\Delta E = 0.19$ eV is the difference between the first 
ionization potentials of oxygen and hydrogen and $T$ is the local 
temperature.

\subsection{Comparison to Observed Reionization History}\label{ssec:reionhist}
A challenge to modeling reionization involves the problem of creating
a high enough ionizing emissivity at early times to match the 
observed optical depth to Thomson scattering in the cosmic microwave
background $\tes$ without overproducing the observed amplitude of the 
EUVB after $z=6$.  Models that assume that a constant 
fraction $\fesc$ of all ionizing photons escape into the IGM can match one, 
but not both of these constraints~\citep{fin11b}.  Observations can be 
reconciled by assuming that $\fesc$ varies with either halo 
mass~\citep{alv12,yaj11} or redshift~\citep{kuh12a,mit13}.  Our fiducial simulation 
uses a time-dependent $\fesc$ to overcome this problem 
(Section~\ref{ssec:sims}).  Here we briefly discuss how well it 
matches observational constraints.

If we assume that helium is singly-ionized with the same neutral fraction
as hydrogen for $z>3$ and doubly-ionized at lower redshifts, then our 
r9n384wWwRT48d simulation yields an integrated optical depth of 
$\tes=0.071$.  This falls within the observed 68\% confidence interval of 
$0.081\pm0.012$~\citep{hin12}, indicating that reionization is sufficiently
extended\footnote{In~\citet{fin12}, we noted that the 
predicted $\tes$ of 0.071 underproduced the observations reported
in~\citet{kom11}.  The current agreement results from the fact 
that measurements of small-scale anisotropy in the CMB have since
brought the inferred $\tes$ down~\citep{hin12,sto12}.  Considering
broader classes of reionization histories also decreases the 
inferred $\tes$~\citep{pan11}.}.  The 
predicted optical depth in the Lyman-$\alpha$ transition at $z=6$ 
is 2.6.  As before, this is somewhat lower than the observed lower 
limit ($>5$;~\citealt{fan06}), implying that the predicted radiation 
field is slightly too strong.  If true, then our simulations could 
underestimate the abundance of OI absorbers at $z=6$.  However, we 
note that our model is not unique in failing 
to reproduce the weak radiation field observed at $z=6$.  In particular, 
observations suggest that the ionizing emissivity strengthens from $<2.6$ to 
$4.3\pm2.6$ (in units of $\times10^{50}$s$^{-1}$Mpc$^{-3}$) from $z=6$ to 
$z=5$~\citep{kuh12a}; such rapid growth is quite difficult to accommodate 
within a model where $\fesc$ varies smoothly with redshift (see, 
however,~\citealt{alv12}).  For redshifts below $z=6$, we 
use predictions from the r6n256wWwRT16d run, which incorporates the same
physical treatments as the fiducial simulation but subtends a smaller
volume.  At $z=5$, this simulation yields an effective optical depth 
to Lyman-$\alpha$ absorption of $\ta=3.1$, marginally consistent with 
the observed range of 2--3~\citep{fan06}.

In summary, the assumption of an evolving escape fraction allows
our simulations to match the observed $\tes$ while only weakly
conflicting with constraints on the post-reionization EUVB.  
Hence the predicted IGM ionization structure, thermal
history, and the star formation history are plausible starting 
points for studying low-ionization metal absorbers during the 
reionization epoch.  In this work, we will show that they primarily
trace star formation in low-mass halos and use their predicted 
abundance as a new test of the model.

\subsection{The Importance of Self-Shielding}
\begin{figure}
\centerline{
\setlength{\epsfxsize}{0.5\textwidth}
\centerline{\epsfbox{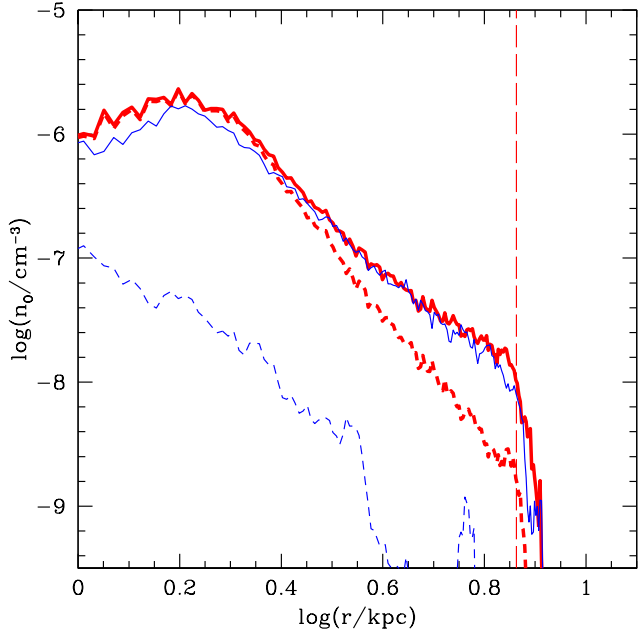}}
}
\caption{The radial density profiles of all oxygen (solid) and
neutral oxygen (dashed) in halos of mass $10^{9.5}\msun$ in simulations 
without (blue) and with (red) self-shielding.  Gas associated with galaxies
has been removed.  The right vertical dashed segment indicates the virial 
radius.  Self-shielding enhances the neutral oxygen abundance significantly 
at all radii.
}
\label{fig:rpOIss}
\end{figure}

Having introduced our simulations, we are now in a position to demonstrate 
the importance of self-shielding.  We compare in Figure~\ref{fig:rpOIss} 
the mean radial density profiles of all oxygen 
(solid) and neutral oxygen (dashed) in simulations without (light blue) 
and with (heavy red) self-shielding (the r6n256wWwRT and r6n256wWwRT16d 
simulations, respectively).  We produce these curves by averaging over 
halos in bins of mass and radius; see Section~\ref{ssec:radprof} for 
details.  The solid curves overlap, indicating that
simulations with similar reionization histories and identical models
for galactic outflows yield similar metal density profiles.  By contrast, 
the light blue dashed curve lies nearly a factor of 10 below the heavy red 
dashed curve, indicating that the neutral oxygen abundance is artifically
underestimated by a factor of $\sim10$ if self-shielding is ignored.  It is
interesting to note that our previous simulations underpredicted the 
observed abundance of OI at $z=6$ by a factor of $\approx15$ 
(Figure 11 of~\citealt{opp09}), independent of whether the ionization state
was modeled using a spatially homogeneous EUVB or a background
dominated by the nearest galaxy.  In that work, the offset was interpreted
as evidence for a partially-neutral universe at $z=6$.  By contrast, 
Figure~\ref{fig:rpOIss} suggests that the disagreement may owe to 
the absence of self-shielding in that work.  If so, then OI observations 
may indeed be consistent with a reionized universe at $z=6$, with the observed 
systems arising entirely in optically thick regions such as galaxies.  Our
new simulations enable us to explore this possibility.

\section{Metal Enrichment and Ionization}\label{sec:ZxHI}

\subsection{The Competition Between Enrichment and Reionization}
\begin{figure}
\centerline{
\setlength{\epsfxsize}{0.5\textwidth}
\centerline{\epsfbox{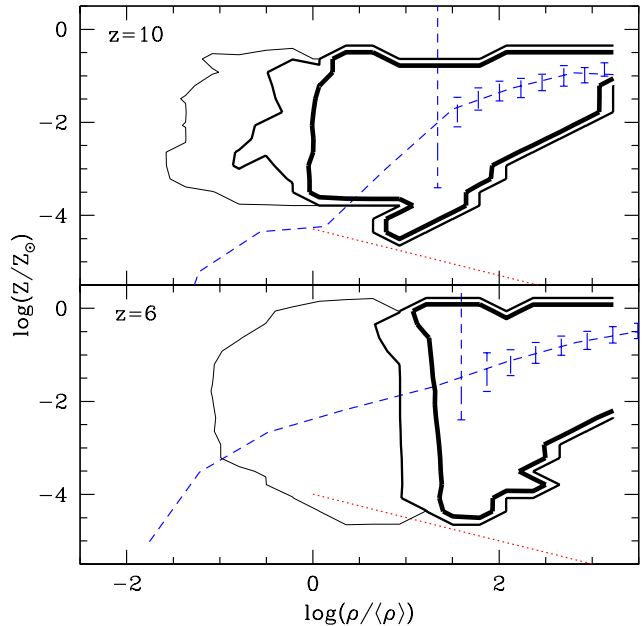}}
}
\caption{The relationship between metallicity, overdensity, neutral fraction, and
reionization at $z=10$ (upper panel) and $z=6$ (lower panel).  Blue dashed curves 
show the mean trend of metallicity versus overdensity while blue dashed error bars
enclose the middle 50\% wherever the median is nonzero.  Light, medium, and heavy 
black contours represent neutral hydrogen fractions of $10^{-5}$, $10^{-2}$, and 
0.5, respectively.  The volume-averaged neutral fractions at $z=6$ and $z=10$ are
0.003 and 0.83, respectively.  Red dotted curves indicate the minimum metal mass
fraction to produce an observable absorber for a hydrostatically-bound region 
at $10^4$K.
}
\label{fig:rhoZxHI}
\end{figure}

Early interest in low-ionization metal absorbers centered on the 
possibility that the diffuse IGM could be enriched before it was 
reionized~\citep{oh02,fur03}.  The question of whether this works 
can be distilled to a competition between the growth of enriched 
regions and the growth of ionized regions.  If galaxies reionize 
their environments more quickly than they enrich them, then 
OI absorption will be dominated by self-shielded clumps rather than 
by low-density regions that have not yet been reionized.  On the other
hand, if galactic outflows enrich the diffuse IGM (that is, regions 
with overdensity $\rho/\langle\rho\rangle < 10$) very quickly, then 
there may be a substantial reservoir of neutral metals that can be 
observed in absorption prior to the completion of reionization.
This idea seems unlikely at a glance because a galaxy's ionization front 
ought to grow more rapidly than its metal pollution front.  However, 
ionizing sources are not necessarily time-steady, and if star formation 
is bursty then 
the IGM surrounding a galaxy can recombine once its OB stars evolve off the 
main sequence.  The metals ejected into the IGM are permanent, however, 
and could become visible in low-ionization transitions~\citep{oh02}.

In order to motivate a detailed study of how this competition unfolds, we 
show in Figure~\ref{fig:rhoZxHI} the relationship between overdensity, 
metallicity, and neutral hydrogen fraction before and after the completion
of reionization.  The blue dashed curves show the mass-weighted mean metal 
mass fraction as a function of overdensity.  
As was seen in Figure 4 of~\citet{opp09}, the mean metallicity grows 
significantly in regions that are moderately overdense 
($\rho/\langle\rho\rangle < 100$) while in denser regions it rapidly 
reaches an equilibrium value that is driven by self-regulated star-forming 
regions~\citep{fin08}.  Importantly, outflows give rise to a reservoir of 
enriched gas at overdensities of 0.01--1 even at $z=10$.  The red dotted 
curves show the minimum metal mass fraction for neutral regions in 
hydrostatic equilibrium at a temperature of $10^4$ K to produce an OI column 
greater than $10^{14}$cm$^{-2}$ as a function of overdensity.  Comparing the 
red dotted and blue dashed curves indicates that overdense regions would 
produce observable absorption if they were homogeneously enriched to the
mean level and neutral.

In order to ask whether the enriched regions could be neutral, we use contours 
to show the neutral hydrogen fraction as a function of density and metallicity.  
The heaviest or innermost contours illustrate the phase space where the 
neutral hydrogen fraction is $\geq50\%$, hence they mark the 
transition from diffuse, ionized gas to condensed, neutral gas.  The low-density
limit of this region lies near the mean density at $z=10$, implying that much of
the metal mass that is expelled into the IGM may remain neutral.  Even at $z=6$,
the bulk of the gas in the Lyman alpha forest ($\Delta\sim10$) is on average
neutral and enriched, implying the presence of a substantial forest of 
low-ionization metal absorbers.

Figure~\ref{fig:rhoZxHI} seems to support the use of OI to probe the progress 
of reionization, but this could be misleading.  The crucial question is whether 
the enriched regions are neutral and vice-versa.  For example, a small 
population of enriched, ionized lumps could drive up the mean metallicity 
without suppressing the mean neutral fraction.  To amplify this possibility,
we use blue dashed error bars enclose the middle 50\% of metallicities 
wherever the median metallicity is nonzero.  They agree with the mean for 
overdensities above $\approx30$, but at lower densities the median vanishes, 
indicating that the mean is driven by a small set of enriched regions.  
The need for detailed study of the IGM phase structure is further
emphasized by observational evidence that metals mix quite poorly with 
the ambient IGM~\citep{sch07}.  If the ionization state is similarly 
inhomogeneous, then the heavy averaging inherent in Figure~\ref{fig:rhoZxHI} 
could be quite misleading.  Our simulations model the inhomogeneous 
ionization and metallicity fields directly (subject to resolution
limitations as described in Section~\ref{ssec:sims}), allowing 
us to address these questions.

\begin{figure*}
\centerline{
\setlength{\epsfxsize}{0.8\textwidth}
\centerline{\epsfbox{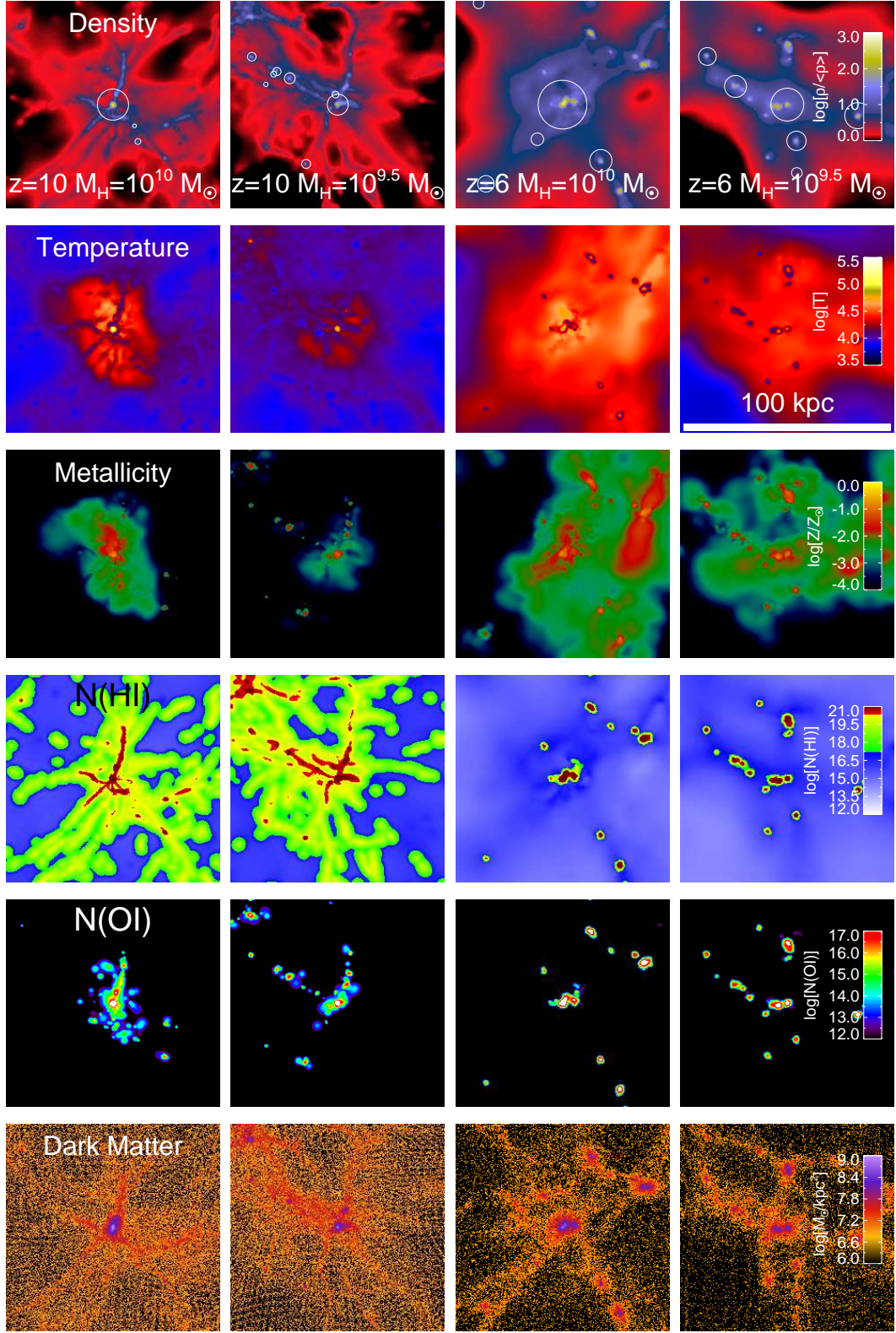}}
}
\caption{Maps of gas density, temperature, metallicity, HI column, OI column,
and dark matter density for two halo masses at $z=10$ and $z=6$.  Each
panel spans 100 proper kpc, and the circles indicate the virial radii of the 
parent halos.  At $z=10$, the weak EUVB leaves an
abundant population of Lyman limit systems, but only those that lie near
halos are associated with a significant OI column.  By $z=6$, OI absorbers 
have retreated well into the central halo's virial radius and are of 
generally higher column density.  
}
\label{fig:rhomap}
\end{figure*}

In order to gain intuition into where OI absorbers live with respect to 
dark matter halos and LLSs, we show in 
Figure~\ref{fig:rhomap} maps of (top to bottom) gas density, 
temperature, metallicity, HI column, and OI column for four 
different dark matter halos at two different redshifts.  The left two
columns show how, at $z=10$, much of the volume is filled with neutral hydrogen
as expected for a universe that is only 50\% ionized.  Near halos, this 
enriched gas produces OI columns stronger than $10^{14}$cm$^{-2}$ well 
outside of the virial radius.  By $z=6$ the gas around similarly massive 
systems (right two columns) is even more enriched, but by now the 
ionization fronts have penetrated deeper into the halo, ionizing much of the 
diffuse gas that would have been visible as low-ionization absorbers at 
$z=10$.  Countering this trend is the growing abundance of satellite halos, 
the cores of which are neutral and enriched.  As a result, low-ionization 
absorbers are common around halos at both $z=10$ and $z=6$.

\begin{figure}
\centerline{
\setlength{\epsfxsize}{0.5\textwidth}
\centerline{\epsfbox{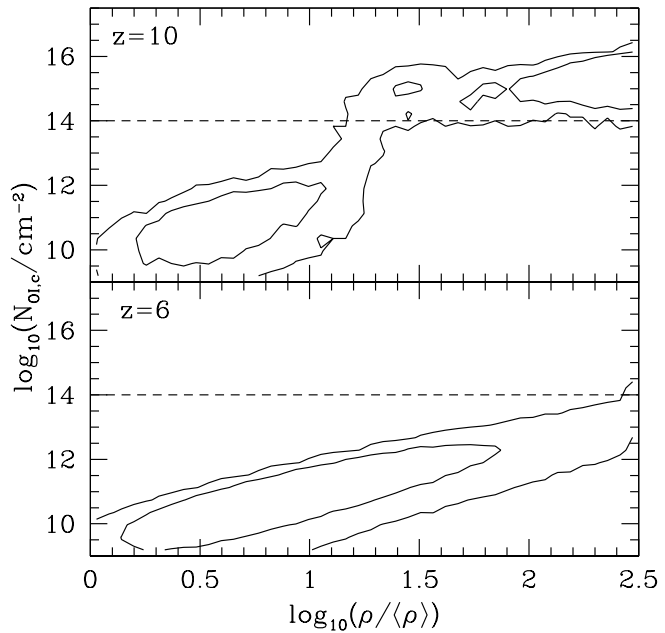}}
}
\caption{The characteristic column density for OI absorption as a function of
overdensity in regions with nonzero metallicity.  
Contours enclose 67\% and 99\% of gas particles at $z=10$ (top) 
and $z=6$ (bottom).  The dashed line indicates the 50\% completeness 
limit~\citep{bec11}.  Regions with overdensities of less than 10 are never 
visible in absorption for $z<10$.
}
\label{fig:rhoNOIc}
\end{figure}

Figure~\ref{fig:rhomap} strongly suggests that OI absorbers trace 
enriched gas within dark matter halos rather than the diffuse IGM.  A 
more quantitative way to ask which regions contain gas that is both enriched 
and neutral enough to yield observable absorption is to compute the 
characteristic column density as a function of density.  If a parcel 
of gas is in hydrostatic equilibrium, then its characteristic OI 
column density $N_{\mathrm{OI},c}$ is
\begin{eqnarray}\label{eqn:NOIc}
N_{\mathrm{OI},c} \equiv L_J \rho_b \frac{Z_O}{m_O} \frac{n_{OI}}{n_O}
\end{eqnarray}
where $L_J$ is the Jeans length, $\rho_b$ is the mass density in baryons, 
$Z_O$ is the mass fraction in oxygen, $m_O$ is the mass of an oxygen 
atom, and $n_{OI}/n_O$ is the neutral oxygen fraction (see Equations 
3--4 of~\citealt{sch01}).  We compute the characteristic column density 
for each overdense particle using the local density,
temperature, metallicity, and ionization state, and show the resulting 
trends at two representative redshifts in Figure~\ref{fig:rhoNOIc}.  
The dashed horizontal shows the current 50\% observational completeness 
limit for selecting absorbers in OI~\citep{bec11}.  Gas at the mean 
density ($\rho/\langle\rho\rangle\sim1$) is ionized by the
nascent EUVB even at $z=10$, hence it does not produce observable
OI absorption.  While we cannot apply Equation~\ref{eqn:NOIc} to 
underdense gas because it is not expected to be in hydrostatic
equilibrium~\citep{sch01}, the trend in Figure~\ref{fig:rhoNOIc} 
strongly suggests that it does not produce visible absorption either.
At higher densities, the threshold for gas to be optically 
thick and hence neutral grows from $\approx20$ at $z=10$ to $>300$ at 
$z=6$.  Given that gas with overdensity greater than 10 is predicted 
to be enriched (Figure~\ref{fig:rhoZxHI}), the evolving threshold for 
it to be optically thick is also the threshold for it to produce 
visible OI absorption.

In summary, our simulations predict that ionization fronts precede 
metal pollution fronts, and that regions, once ionized, remain ionized.
This owes partially to the fact that hydrogen-cooling halos produce
stars steadily until their environments are reionized (note 
that~\citealt{wis08} find that star formation becomes a steady-state
process in pre-reionization halos more massive than $10^7\msun$, an order of magnitude
below our resolution limit) and partially to the clustered nature of 
galaxy formation, although a detailed analysis of the relative roles 
of these factors is currently impossible owing to our small volumes.
Consequently, diffuse gas does not produce observable absorption in 
low-ionization transitions.  For the rest of this work, we will therefore
focus on low-ionization metal absorption that occurs within dark 
matter halos.

\subsection{Radial Profiles}\label{ssec:radprof}

In this Section, we explore how the radial density profiles of gas, total 
metals, and neutral metals vary with mass and redshift.  We will consider 
halos that are both more and less massive than $10^9\msun$ because this 
marks the approximate threshold above which halos can accrete gas even 
in the presence of an EUVB.  For consistency 
with~\citet{fin11b}, we will refer to the lower-mass halos as 
``photosensitive" and the more massive halos as ``photoresistant".

\begin{figure}
\centerline{
\setlength{\epsfxsize}{0.5\textwidth}
\centerline{\epsfbox{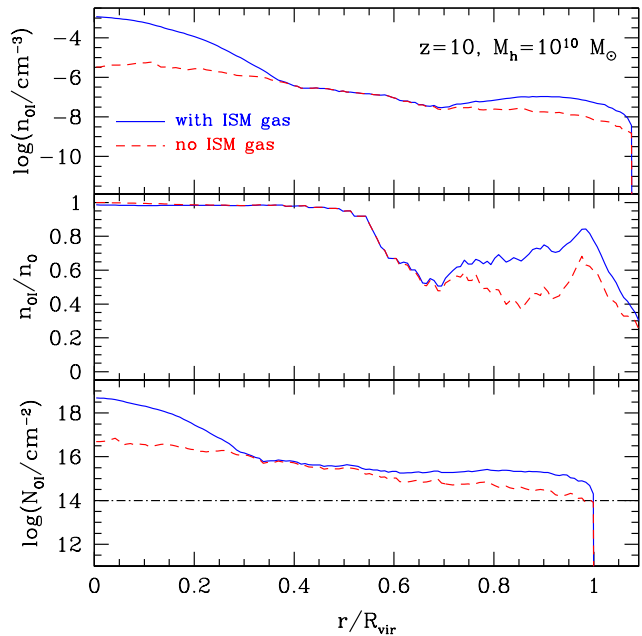}}
}
\caption{Sample profiles of OI density $n_{OI}$, neutral oxygen 
fraction $X_{OI}$, and neutral oxygen column density $\NOI$ as a function 
of radius for the $10^{10}\msun$ central halo at $z=10$ shown in the left 
column of Figure~\ref{fig:rhomap}.  The solid blue
profiles include both interstellar and intergalactic gas, while the dashed
red profiles exclude all gas that is bound within resolved galaxies.  
The $n_{OI}$ and $X_{OI}$ profiles are both smoothed with a 1-kpc boxcar 
filter.  The black dot-dashed curve in the bottom panel indicates current
observational limits~\citep{bec11}.  We do not trace the profiles 
beyond a virial radius, hence they vanish there artificially.  Excluding 
ISM gas suppresses the OI column at small radii owing to the central galaxy 
and at large radii owing to satellites, but on the whole the halo remains 
observable out to the virial radius.
}
\label{fig:rhoSamp}
\end{figure}

We compute radial density profiles by stacking halos in bins of mass and 
averaging within each radial bin.  By computing the density of OI within 
each shell directly (rather than computing the oxygen density and neutral 
fractions and multiplying them), we preserve small-scale inhomogeneities 
in the metallicity and enrichment fields.

As a demonstration of how our spherically-averaged radial profiling works, 
we show in Figure~\ref{fig:rhoSamp} the density, neutral fraction, and 
column density profiles for our most massive halo at $z=10$ (left colum 
in Figure~\ref{fig:rhomap}).  The solid blue curve
shows that the halo possesses an enriched neutral core that is 
associated with OI column densities above $10^{16}$cm$^{-2}$ out to at 
least 0.2 virial radii ($\rvir = 6.6$kpc; bottom panel).  This is 
dominated by star-forming gas in the central galaxy.  Outside of this 
core there is an enriched, partially-neutral reservoir that 
generates observable column densities ($\NOI>10^{14}$cm$^{-2}$; 
the black dot-dashed line in the bottom panel) out to the virial radius.

Our approach works well if the gas is distributed spherically-symmetrically, 
but it breaks down if the majority of a halo's gas is bound into a small 
number of satellite systems because the geometric cross section for a sightline 
to intersect a satellite is smaller (and the associated gas column higher) than 
if the satellite's gas were distributed in a shell.  Additionally, the fact 
that our simulations neglect ionizations owing to the local radiation field 
means that the abundance of neutral oxygen within galaxies could be 
overestimated (we will return to this point in section~\ref{sec:discuss}).  
In order to mitigate these problems, we use
{\sc SKID}\footnote{\url{http://www-hpcc.astro.washington.edu/tools/skid.html}}
to identify and remove all gas that is associated with galaxies
before computing density profiles.  The dashed red curve shows the same 
density profile as the solid blue curve, but without galaxy gas.  This 
step suppresses the density of neutral gas significantly near the halo's 
core, but at larger radii the difference is slight because the gas in
resolved satellites is subdominant to the combined contributions of 
unresolved satellites and the circumgalactic medium (CGM).

Note that the column densities in the bottom panel are notional because they
are derived from spherically-averaged profiles.  In the second part of 
this work, we will relax the assumption of spherical symmetry and use a 
ray-casting approach to compute the geometric absorption cross section, 
enabling a more accurate comparison with observations.

\begin{figure}
\centerline{
\setlength{\epsfxsize}{0.5\textwidth}
\centerline{\epsfbox{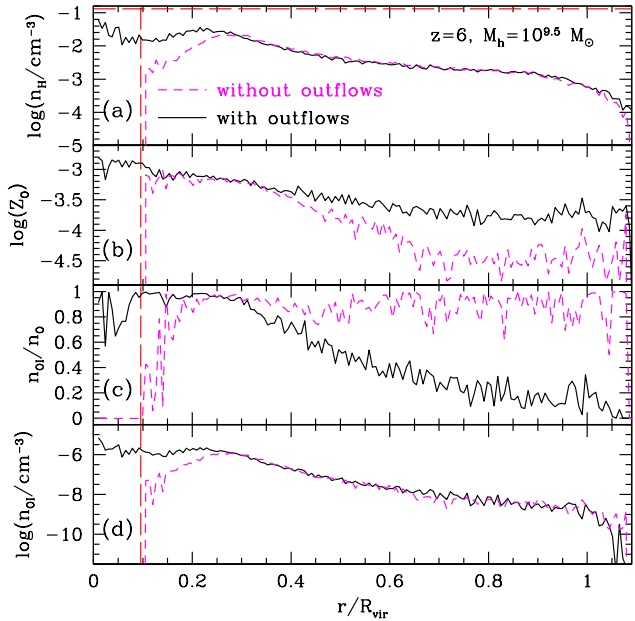}}
}
\caption{The radial profiles of hydrogen density, oxygen mass fraction, 
and neutral fraction, and neutral oxygen density in a $10^{9.5}\msun$ 
halo at $z=6$ in simulations without outflows (magenta dashed) and 
with outflows (black solid).  The virial radius is 7.3 kpc.  The 
red horizontal long-dashed line in the top panel indicates the threshold 
density for forming stars.  The red vertical long-dashed line
indicates the gravitational softening length.  Outflows dominate the
CGM at small radii and generate an atmosphere of ionized, enriched gas 
at large radii.  They do not enhance the geometric cross section for 
absorption in low-ionization transitions.
}
\label{fig:rpoutflows}
\end{figure}

Having demonstrated how we compute spherically-averaged profiles, we 
now ask how outflows impact the CGM.  We show 
in Figure~\ref{fig:rpoutflows} the radial density
profiles of gas and metals in $10^{9.5}\msun$ halos in simulations 
without (dashed magenta) and with (solid black) galactic outflows.  
Panel (a) compares the gas densities.  The profiles flatten below
one kpc because gas at these radii is dense enough to support star
formation, which suppresses the gas density.  Recalling that we have 
removed galactic gas from these profiles, we see that there is no 
circumgalactic gas within $0.1\rvir$ unless outflows put it there
because inflows  at these radii collapse quickly onto the 
central galaxy.  At larger radii, the profiles are nearly coincident 
because most of the gas is infalling rather than outflowing.

Panel (b) shows the oxygen metallicity profile.  Near the
central star-forming region (within 2 kpc), outflows give rise to
an enriched atmosphere.  At larger radii, simulations without outflows
still suggest an enriched CGM.  However, outflows clearly boost the 
mean metallicity beyond  $0.2\rvir$~\citep[see also][]{opp09}.

We show the neutral oxygen fraction in panel (c).  
Nearly all of the CGM's metals are neutral in the absence of 
outflows.  These metals could correspond either to star-forming 
gas in satellite halos that are too small to be identified and 
removed by our group finder, or to moderately-enriched inflowing 
streams; simulations with higher resolution would be required to 
distinguish between these possibilities.  By contrast, the neutral 
metal fraction drops at large radii in simulations with outflows.  This 
does not owe to differences in the EUVB because $\fesc$ is tuned 
separately for each simulation to produce similar EUVBs by $z=5$.
Instead, it indicates that outflows tend to be highly ionized.
A detailed analysis of the thermal structure of outflows is beyond
the scope of the present work, but for reference we note that, for
gas particles that have recently been ejected at $z=7$, our model
predicts a median density of 0.4 times the mean baryon density
and a median temperature of 30,000 K.  For such gas, the 
recombination time exceeds a Hubble time, hence it is expected 
to be largely ionized by $z=7$.

The product of the curves in the top three panels is proportional to 
the neutral oxygen density, which we show in panel (d).  This
panel confirms that metals that are ejected in outflows are generally
ionized and do not enhance the probability that the host halo will
be observable as a low-ionization metal absorber.  They must instead
be sought using high-ionization transitions such 
as CIV~\citep{opp06,bor13} or OVI~\citep{tum11}.  Note that this
conclusion is not necesssarily general.  For example,~\citet{for13} 
has shown that outflows enhance the abundance of MgII absorbers 
around $10^{12}\msun$ halos at low redshifts (their Figure 14).

\begin{figure}
\centerline{
\setlength{\epsfxsize}{0.5\textwidth}
\centerline{\epsfbox{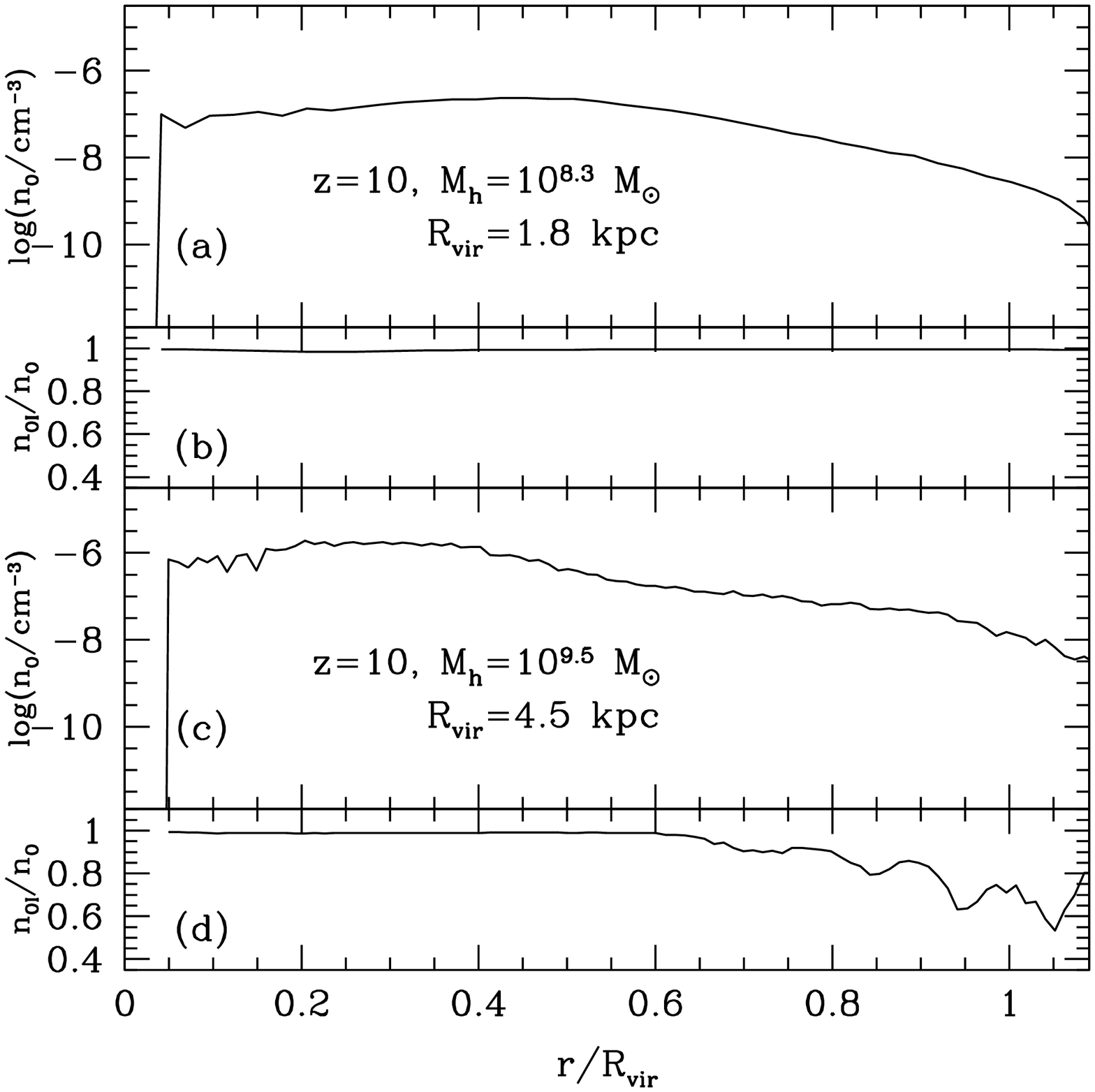}}
}
\caption{
The radial profiles of oxygen density 
in halos of mass $\log_{10}(M_h/M_\odot)=8.3$ (panel a) 
and 9.5 (panel c) at $z=10$ in our fiducial simulation
Panels (b) and (d) show the corresponding neutral
fractions.  Profiles are smoothed with a 0.3-kpc boxcar 
filter for clarity.  At $z=10$, CGM metals are completely
neutral in photosensitive halos and mostly neutral in
photoresistant halos.
}
\label{fig:rpmassz10}
\centerline{
\setlength{\epsfxsize}{0.5\textwidth}
\centerline{\epsfbox{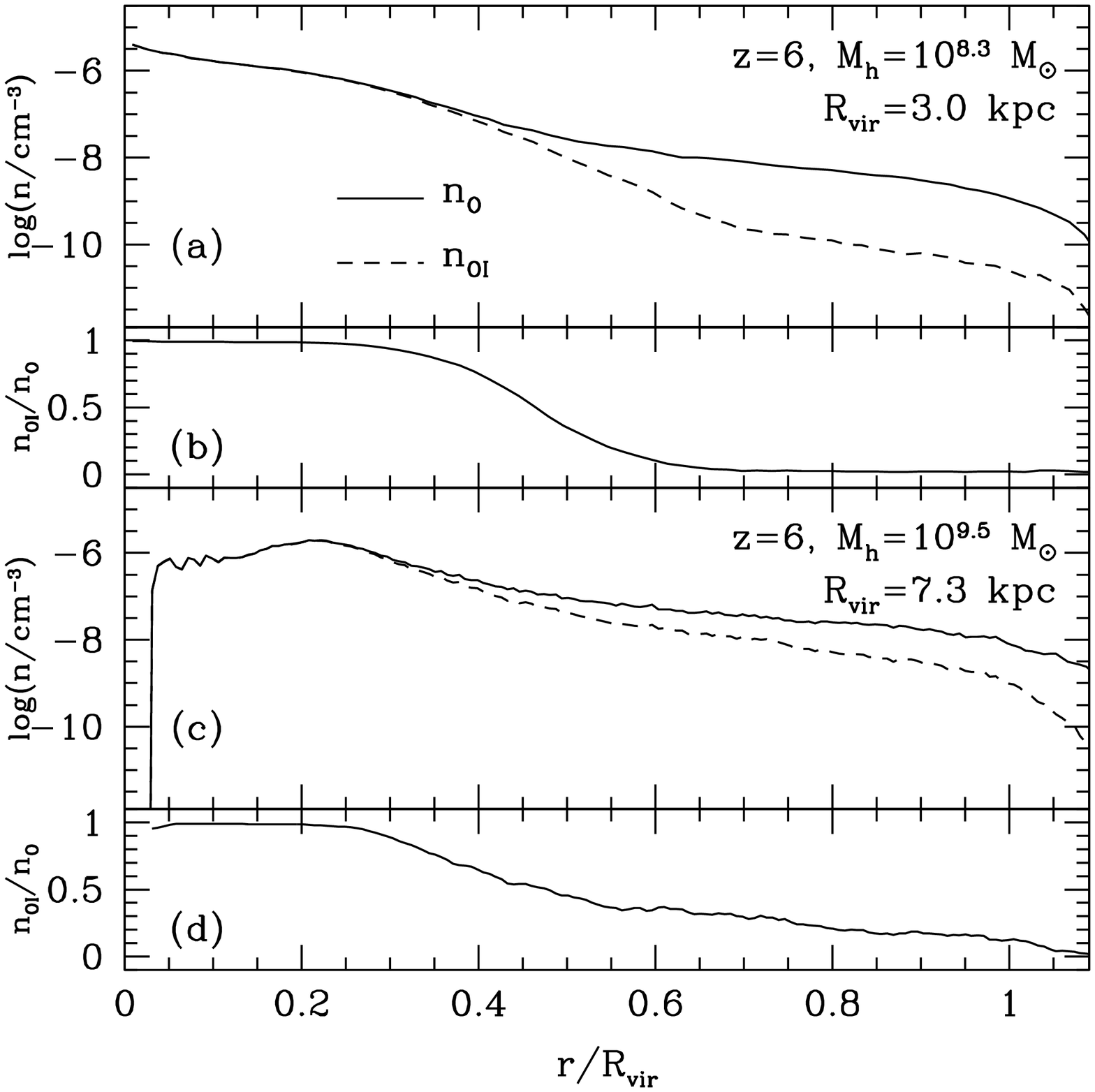}}
}
\caption{The same as Figure~\ref{fig:rpmassz10} but at $z=6$.  We also
distinguish total and neutral oxygen density in panels (a) and (c) as
indicated.  Once reionization completes, the EUVB penetrates to roughly
$0.5\rvir$ in both photosensitive and photoresistant halos.  A tail
of partially-neutral gas extends to the virial radius in the photoresistant
halos, suggesting that they could dominate OI absorption statistics once 
reionization is complete.
}
\label{fig:rpmassz6}
\end{figure}

In Figure~\ref{fig:rpmassz10}, we evaluate how the OI density profile 
varies with mass prior to the completion of reionization.  Examining 
the photosensitive halos first (panel a), we find that the central 
star-forming region ($<0.5\rvir$) contains a significant reservoir of 
metals because these halos are massive enough to cool their gas and 
form stars.  Furthermore, panel (b) shows that their metals remain 
completely neutral out to the virial radius because they inhabit preferentially 
underdense regions where the EUVB remains weak at $z=10$.  We will 
show in Sections~\ref{ssec:sigma}--\ref{ssec:dNdXdM} that these halos 
have a geometric cross section to absorption that is not small 
compared to the halo cross section and that, consequently, they 
dominate OI absorption statistics prior to the completion of 
reionization.

Turning to photoresistant halos, we see that the density of metals
is 1--2 orders of magnitude higher than in the photosensitive halos owing
to their higher star formation efficiencies.  The neutral fraction drops
below unity outside of roughly $0.6\rvir$ because these halos inhabit
preferentially dense regions where the EUVB takes hold at earlier times. 
Even at the virial radius, however, the neutral fraction exceeds 50\%, 
suggesting that these halos generate high-column absorbers even though 
they are subject to a stronger EUVB.

As the EUVB strengthens and ionization fronts penetrate the CGM, we 
expect the OI density profiles to evolve.  We show in 
Figure~\ref{fig:rpmassz6} how the profiles in the same mass ranges have 
evolved by $z=6$ (note that the virial radii are also larger now).  
Photosensitive halos have grown a 
substantially higher total oxygen density, particularly near their cores
($\leq0.3\rvir$).  These halos are able to continue forming new stars and
metals even at $z=6$ owing to the fact that gas that cooled prior 
to reionization remains bound and star-forming for several dynamical 
times following overlap~\citep{dij04}.  However, the gas is only neutral
within $0.5\rvir$.  Gas at larger radii is completely ionized by the 
EUVB.  In fact, our simulations 
suggest that halos near the hydrogen-cooling limit ($\sim10^8\msun$) are 
evaporated by the EUVB in a process similar to the evaporation of minihalo 
gas~\citep{sha04}.  For both of these reasons, the abundance of neutral 
oxygen at the virial radius of photosensitive halos declines and their 
contribution to low-ionization metal absorbers diminishes as reionization
proceeds.

Photoresistant halos are also more ionized than at $z=10$ (panels c and d),
but the effect is weaker than for photosensitive halos.  In detail, the
fraction of neutral metals drops below 50\% at roughly $0.5\rvir$ in both
cases, but the photoresistant halos are able to retain a significant 
component of neutral gas out to nearly the virial radius.  Moreover,
the total mass of circumgalactic metals around massive halos grows owing 
to continued star formation, metal expulsion, and possibly stripping of 
enriched gas from infalling satellites, as can clearly be seen in 
Figure~\ref{fig:rhomap}.  This means that, although photoresistant halos 
are exposed to a generally stronger EUVB, their denser CGM are able to 
attenuate the ionization fronts and preserve a reservoir of neutral metals 
that extends throughout much of the halo even at $z=6$.

In summary, Figures~\ref{fig:rpmassz10}--\ref{fig:rpmassz6} suggest 
that the overall abundance of OI absorbers is regulated by a competition between 
the halo abundance, which grows in time, and absorption cross section, which
declines for all halo masses as time progresses.  All halos generate an 
enriched CGM down to the hydrogen cooling limit.  The metals remain largely neutral at 
$z=10$ such that photosensitive halos are the predominant source of 
low-ionization metal absorbers prior to reionization.  Nearer the epoch 
of overlap, photosensitive halos are completely ionized at radii larger 
than $0.5\rvir$ whereas photoresistant halos are more than 10\% neutral 
out to the virial radius.  Hence the typical host halo mass of OI 
absorbers increases as reionization proceeds.  We will quantify this 
evolution in Figures~\ref{fig:MfromN} and~\ref{fig:dNdXdM}.

\section{Modeling Observations}\label{sec:obs}
In this section we relate the properties of individual halos to volume-averaged
statistical measurements of OI.  Our analysis follows the approach adopted by 
many previous numerical studies of DLAs~\citep[for example,][]{kat96}.
We begin by computing the geometric cross-section
for halos to be observed in absorption in a way that relaxes the assumption of
spherical symmetry.  We then study how reionization affects the appearance of 
different halos in absorption.  Finally, we apply our cross sections to predict 
the observable number density of absorbers and compare directly to observations.

\subsection{Cross-section for Observability}\label{ssec:sigma}

\begin{figure}
\centerline{
\setlength{\epsfxsize}{0.5\textwidth}
\centerline{\epsfbox{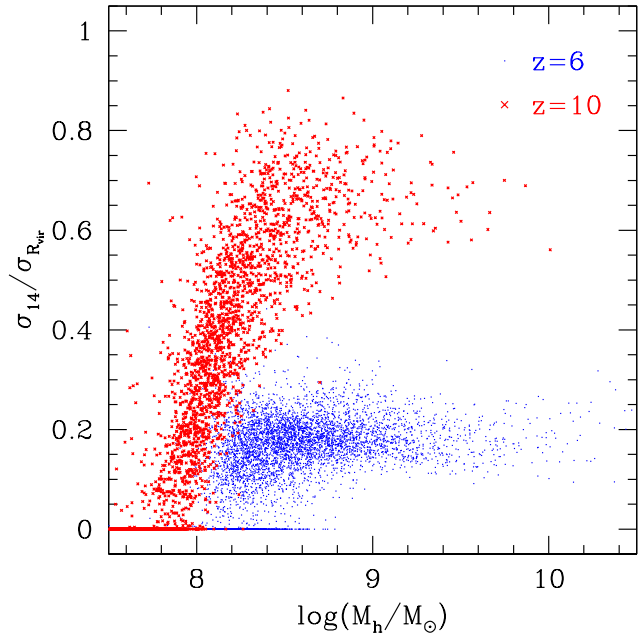}}
}
\caption{The fraction of the area within one virial radius 
$\pi\rvir^2$ that is covered by lines of sight with a neutral oxygen 
column greater than $10^{14}$cm$^{-2}$ as a function of halo mass at 
$z=10$ (red crosses) and $z=6$ (blue points).
}
\label{fig:sigvmass}
\end{figure}

Halos that are more massive at a given redshift or at lower
redshift for a given mass have produced more metals, leading
to a higher cross section.  Similarly, halos with lower mass
at a given redshift should have a lower cross section both because 
they have produced fewer metals and because they are more susceptible 
to an EUVB.  Our simulations allow us to quantify
these effects with minimal assumptions.  We begin by computing the 
geometric cross section for a halo to appear as an OI absorber with a 
column density greater than $10^{14}$cm$^{-2}$, which is the 50\% 
completeness limit reported by~\citet{bec11}.

Computing the cross-sections accurately requires us to relax the assumption 
of spherical symmetry because the OI column density profiles are influenced by 
the filamentary structure of the gas density field (Figure~\ref{fig:rhomap}).
We map each of our halos onto a mesh with cells of width 200 physical 
parsecs including all gas out to twice the virial radius and then count the 
fraction of lines of sight passing within one virial radius for which the
OI column density exceeds $10^{14}$cm$^{-2}$.  We recompute this fraction 
using lines of sight in the x, y, and z-directions and average the three 
results.  Using a finer mesh decreases the cross section while increasing 
the number of lines of sight with high columns, but the effect is weak; we 
have verified that using a mesh with twice this spatial resolution changes 
the cross-sections by $\sim10\%$.  Considering lines of sight that pass
outside of one virial radius would primarily have the effect of picking
up absorption owing to neighboring halos, as can be seen at $z=6$ in 
Figure~\ref{fig:rhomap}.  Incorporating the full 
three-dimensional gas distribution in this way automatically accounts 
for any departures from spherical symmetry.  This means that, whereas 
we excluded gas that is closely 
associated with galaxies in Section~\ref{ssec:radprof} in order not to 
``smear" satellites over spherical shells when computing mean radial 
profiles, we include all halo gas in the analysis throughout the rest of 
this work.

We show in Figure~\ref{fig:sigvmass} how the fraction of
the area within one virial radius that is covered by observable lines of
sight $\sigma_{14}/\sigma_{\rm vir}$ varies with halo mass at $z=10$ (red 
crosses) and $z=6$ (blue points).  Broadly, $\sigma_{14}/\sigma_{\rm vir} < 1$
even at $z=10$.  In detail, halos more massive than $10^8\msun$ are 
generally visible throughout much of the virial radius because the 
EUVB has not yet penetrated deep into the CGM.  Halos 
less massive than $10^8\msun$ show weaker absorption because they 
are not capable of producing stars and metals even in a neutral IGM.  At 
$z=6$, the signature of reionization is obvious.  The EUVB has penetrated 
well into the typical halo, suppressing the covered fraction to 10--30\%.  
The threshold halo mass below which the absorption cross section vanishes
grows from $10^8\msun$ at $z=10$ to roughly 
2--3$\times10^8\msun$ at $z=6$.  This owes to the combined effects of 
photoionization and gas exhaustion on photosensitive 
halos.~~\citet{raz06} used radiation hydrodynamic simulations to find 
that halos less massive than $7\times10^7\msun$ retain the ability to
accrete gas following reionization.  Our simulations indicate a slightly
higher threshold, likely owing to the tendency for outflows to reduce
the gas density near halo cores.  There is 
also a population of halos at both redshifts that produce no 
observable absorption ($\sigma_{14}=0$).  This population extends to 
higher mass at $z=6$ than at $z=10$, indicating that it is not purely 
an artefact of limited mass resolution; instead, it reflects the weak 
star formation efficiencies and optically-thin CGM of photosensitive halos.

\subsection{The Dominant Host Halos}\label{ssec:MfromN}
\begin{figure}
\centerline{
\setlength{\epsfxsize}{0.5\textwidth}
\centerline{\epsfbox{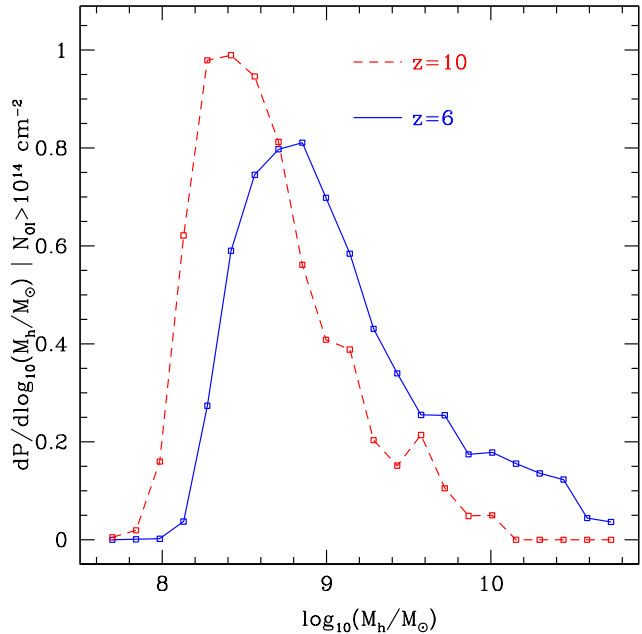}}
}
\caption{The probability density that the host halo of an OI absorber
with column density greater than $10^{14}$cm$^{-2}$ has a given mass
at $z=10$ and $z=6$.  OI absorbers are dominated by halos a factor of
10--100 less massive than the halos that host Lyman break 
galaxies and Lyman-$\alpha$ emitters~\citep{ouc10,mun11}.
}
\label{fig:MfromN}
\end{figure}

As a first application of our cross sections, we may compute the 
most likely host halo mass for absorbers at a given column density.  
We do this by computing the probability density $P(M|N)$ that 
an absorber with column density $\NOI>10^{14}$cm$^{-2}$ is 
hosted by a halo of mass $M < M_h < M+dM$ using Bayes' theorem:
\begin{equation}\label{eqn:PMN}
P(M|N) \propto P(N|M) P(M)
\end{equation}
where $P(N|M)$ is the probability that a line of sight passing within
the virial radius of a halo of mass $M$ encounters a column greater 
than $>10^{14}$cm$^{-2}$ and $P(M)$ is the prior probability of passing
within a virial radius of a halo of mass $M$.  The former is simply
the ratio of the area within which the column exceeds $10^{14}$cm$^{-2}$
to the area within a virial radius (that is, 
$\sigma_{14}/\sigma_{\rm vir}$), and the latter is the fraction of
halos in this mass range weighted by the area within a virial radius.
We show this probability density at $z=6$ and $z=10$ in 
Figure~\ref{fig:MfromN}.

At $z=10$, the distribution of halo masses that can host an
observable system is weighted toward the hydrogen cooling limit partly 
because the enriched CGM in such halos remain mostly neutral,
and partly because more massive halos are not yet abundant enough to
compete.  By $z=6$, the peak of the probability density function has 
shifted to higher mass by a factor of 2--3 because photosensitive halos lose 
their gas while photoresistant halos begin to assemble in force.
Still, however, the characteristic host halo's mass lies within the
range that is sensitive to photoionization heating~\citep{fin11b}.
This suggests that, at any redshift, low-ionization metal absorbers
probe the lowest-mass halos that retain the ability to form stars.

How do the host halos of OI absorbers compare with the host halos
of galaxies that are selected in emission?~~\citet{mun11} used a 
detailed comparison
between an analytic model and observations of Lyman break galaxies 
at $z=7$--8 to show that current observations likely 
do not probe below a halo mass of $\sim10^{10}\msun$. 
Similarly,~\citet{ouc10} have used
clustering observations to infer that Lyman-$\alpha$ emitters live
in halos with masses between $10^{10}$--$10^{11}\msun$.  Hence
absorption-selected samples trace star formation in halos that are 
10--100 times less massive than the halos that host emission-selected
samples.  This supports the suggestion by~\citet{bec11} that studies 
in absorption offer more direct insight into the nature of the systems 
whose ionizing flux may have driven hydrogen 
reionization~\citep[see, for example][]{yan04,alv12,rob13}.

Figure~\ref{fig:MfromN} also gives insight into what would be required
to observe the host galaxies of OI absorbers in emission.  The
typical OI absorber at $z=6$ lives in a $10^9\msun$ halo.  Our models 
predict that the mean star formation rate of such halos is 
0.009$\smyr$~\citep{fin11b}.  Assuming that the ratio of luminosity
to SFR is $2\times10^{28}$ ergs s$^{-1}$ Hz$^{-1} (\smyr)^{-1}$~\citep[][note that this
includes an estimate for dust extinction]{fin11a}, this 
corresponds to a rest-frame ultraviolet absolute magnitude of -14.  
This is roughly three magnitudes fainter than has been 
achieved at $z\geq6$ with the Hubble Space Telescope~\citep{bou12}, 
and slightly fainter than will be achieved with the James Webb 
Space Telescope.

As a caveat to Figure~\ref{fig:MfromN}, we note that the typical 
host halo mass at $z=6$ may be underestimated because the most massive 
halos are undersampled by our small simulation volume.  In 
Figure~\ref{fig:dNdXdM}, we will use an analytic fit to our results to 
extrapolate to higher masses and confirm that photosensitive halos still 
dominate.

\subsection{The Contribution of Halos of Different Masses}\label{ssec:dNdXdM}
\subsubsection{Median Cross Section Versus Mass}
\begin{figure}
\centerline{
\setlength{\epsfxsize}{0.5\textwidth}
\centerline{\epsfbox{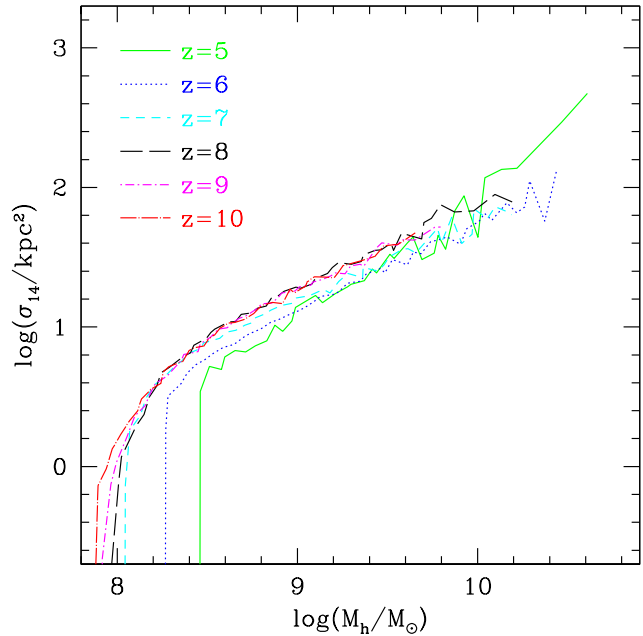}}
}
\caption{
The median cross section for appearing as an absorber with an OI column 
density greater than $10^{14}$cm$^{-2}$ as a function of halo 
mass and redshift.  The $z=5$ curve comes from our r6n256wWwRT16d 
simulation whereas the others all come from the r9n384wWwRT48d run.  
The minimum mass that can produce observable absorption and the 
proper cross section of massive halos both increase with declining 
redshift.
}
\label{fig:sigvMzconst}
\end{figure}

By how much does the cross section shrink from $z=10\rightarrow6$?
Figure~\ref{fig:sigvmass} shows that the covered fraction declines
at constant mass, but the x-axis is in units of virial radii.  In
practice, it is convenient to quantify the absorber abundance in 
terms of the number per absorption path length 
(section~\ref{ssec:dldM}), which in turn depends 
on the cross section in proper units.  To this end, we show in 
Figure~\ref{fig:sigvMzconst} how the median cross section for 
observability $\sigma_{14}$ in proper kpc$^2$ depends on halo mass 
at 6 redshifts.  The 
median trend evolves in three distinct ways.  First, we still see a 
low-mass cutoff that grows owing to the gradual encroachment of ionization 
fronts into photosensitive halos.  As before, the cutoff evolves from 
$<10^8\msun$ at $z=10$ to a few $\times10^8\msun$ by $z=6$.  Second, for 
halos with virial mass $<10^{10}\msun$, the growth of the EUVB 
dominates over the impact of continuing metal enrichment with the result that 
the cross section at a given halo mass shrinks.  This is the signature of 
reionization: as observations probe higher 
redshifts, the EUVB weakens, halos are more neutral, and absorption 
shifts from high-ionization transitions in photoresistant halos to 
low-ionization transitions in photosensitive halos.  This is consistent 
with the observation that the abundance of CIV absorbers declines at 
$z>6$ while the abundance of low-ionization systems does not~\citep{bec11}.
The shift to lower host halo masses may manifest as a decline in
the characteristic velocity width as observations push past $z=5$; we will
explore this possibility in future work.
Finally, the cross section for photoresistant halos grows following $z=6$ 
(that is, the solid green curve lies above the dotted blue one for $z=5$ 
and $M_h/\msun>10^{9.5}$).  This evolution is a direct response to our 
strongly redshift-dependent ionizing escape fraction (Section~\ref{ssec:sims}): 
If $\fesc$ declines more rapidly than the star formation rate density 
increases, then the EUVB amplitude declines.  As it does so, ionization 
fronts recede to larger halocentric radii, leaving more of the CGM neutral 
(since the recombination time remains much shorter than the Hubble time in 
moderately overdense gas at $z=6$).  Whether this behavior is real depends 
on the true evolution of the EUVB.  Current observations of the 
Lyman-$\alpha$ forest suggest that it strengthens 
dramatically from $z=6$ to $z=5$~\citep{bol07,kuh12a} whereas it declines 
in our simulation, hence it may be no more than an artefact of our simple 
parameterization for $\fesc$.  It would be interesting to study how the cross
section varies in a model that assumes a mass-dependent $\fesc$ as such models
may be able to reproduce the observed evolution of the EUVB more 
faithfully (for example,~\citealt{alv12,yaj11}).

The evolution with redshift at the low-mass end may be compared with the 
finding by~\citet{bec11} that the number density of low-ionization absorbers 
does not evolve strongly for $z>3$.  In that work it was proposed that, at 
higher redshift, either the typical halo mass of absorbers decreases or the 
cross section for halos to appear as low-ionization absorbers increases.  
Figure~\ref{fig:sigvMzconst} supports the idea that photosensitive halos, 
which are 
the predominant hosts of OI absorbers, do indeed have larger cross section
in the presence of a weaker EUVB, in qualitative agreement with this scenario.

The trends in Figure~\ref{fig:sigvMzconst} are well fit by a broken 
power law.  We have performed a least-squares fit using the 
form $\log_{10}(\sigma_{14}) = a + b \log_{10}(M_h/\msun)$, where 
$\sigma_{14}$ 
is in proper kpc$^2$ and the fit parameters ($a$,$b$) change from 
($a_l,b_l$) to ($a_h,b_h$) at the cutoff mass 
$\log_{10}(M_{h,c}/\msun)$.  Table~\ref{table:fits} shows the 
resulting fits.  The predicted power-law slope lies in the range
0.7--0.9, slightly steeper than what would be expected if the
cross section for absorption were a constant fraction of each 
halo's virial cross section ($M^{2/3}$).  This indicates that
feedback preferentially suppresses the OI abundance of low-mass 
systems.  These slopes are consistent with the scalings that are
found for absorption by neutral hydrogen (for 
example,~\citet{nag04}), suggesting a physical correspondence between
systems that are selected in OI and HI.

\subsubsection{The Significance of Halos at Different Masses}\label{ssec:dldM}
It is convenient to quantify the number density of absorbers in terms
of the number per absorption path length $l = dN/dX$, where the path
length element $dX$ is defined in such a way that $l$ is constant if 
the comoving number density and proper cross section of the absorbers 
do not evolve~\citep{bah69,gar97}:
\begin{equation}
dX\equiv(1+z)^2\frac{H_0}{H(z)}dz
\end{equation}
The differential number density of absorbers per absorption path 
length per halo mass $M$ owing to absorbers with proper cross 
section $\sigma$ is:
\begin{equation}\label{eqn:dldM}
\frac{dl}{dM} = \frac{c}{H_0}\sigma\frac{dn}{dM},
\end{equation}
where $dn/dM$ is the dark matter halo mass function.

\begin{figure}
\centerline{
\setlength{\epsfxsize}{0.5\textwidth}
\centerline{\epsfbox{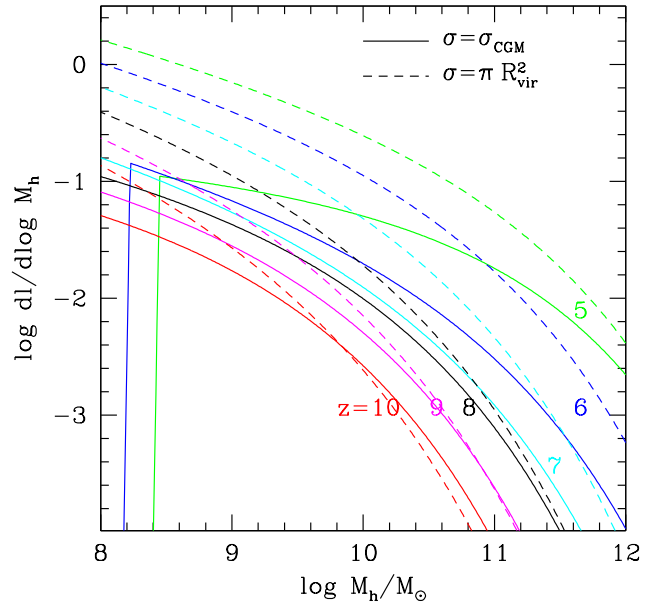}}
}
\caption{The differential contribution to the abundance of OI absorbers
with column density greater than $10^{14}$cm$^{-2}$ as a function of
mass and redshift.  From bottom to top, curves correspond to $z=10,9,8,7,6$
and 5.  Solid contours result from folding the predicted cross sections
from the simulation directly into Equation~\ref{eqn:dldM} while dashed
contours assume that the maximum observable radius is simply the virial
radius.
}
\label{fig:dNdXdM}
\end{figure}

In order to explore how halos of different masses contribute to the
total abundance of observable (column density above $10^{14}$cm$^{-2}$) 
OI absorbers, we combine the analytical fits to our predicted cross 
sections in Table~\ref{table:fits} with the~\citet{she99} halo mass 
function using Equation~\ref{eqn:dldM} and show the resulting 
differential number counts from $z=10\rightarrow5$ using solid curves in 
Figure~\ref{fig:dNdXdM}.  We consider only halos more massive than 
$10^8\msun$ as lower-mass halos tend to be unobservable 
(Figure~\ref{fig:sigvMzconst}).  For comparison, we also show 
the predicted abundance under the assumption that each halo appears 
as an OI absorber out to its virial radius (dashed curves); we will refer 
to this as the ``$\rvir$ model".  Note that Figure~\ref{fig:dNdXdM} is
qualitatively similar to Figure~\ref{fig:MfromN}.  The primary 
difference is that, while Figure~\ref{fig:MfromN} takes the full 
distribution of cross section as a function of halo mass into account, 
Figure~\ref{fig:dNdXdM} extrapolates to higher masses than can be
explored in our simulations' limited cosmological volumes.

The predicted abundance varies slowly from $z=10$ (lowest red 
curve) to $z=5$ (highest green curve) despite the onset of reionization 
at $z\sim10$.  At a given redshift, however, there is a cutoff 
mass (given by the final column in Table~\ref{table:fits}).  Below
this mass, the fractional contribution per unit halo mass vanishes, 
indicating that the rapid decline in cross section toward 
low masses cannot be made up by the increasing halo abundance.  Above 
the cutoff mass, the curves approach the $\rvir$ model because more
massive halos are visible out to a larger fraction of the virial radius.
However, the massive halos do not dominate absorbers by number because
they are too rare.  This confirms the conclusion from 
Figure~\ref{fig:MfromN} that photosensitive halos dominate observations
even when the limitations of our small cosmological volume are 
corrected for.

\subsection{Comparing to the Observed Number Density}\label{ssec:dNdX}
\begin{figure*}
\centerline{
\setlength{\epsfxsize}{0.9\textwidth}
\centerline{\epsfbox{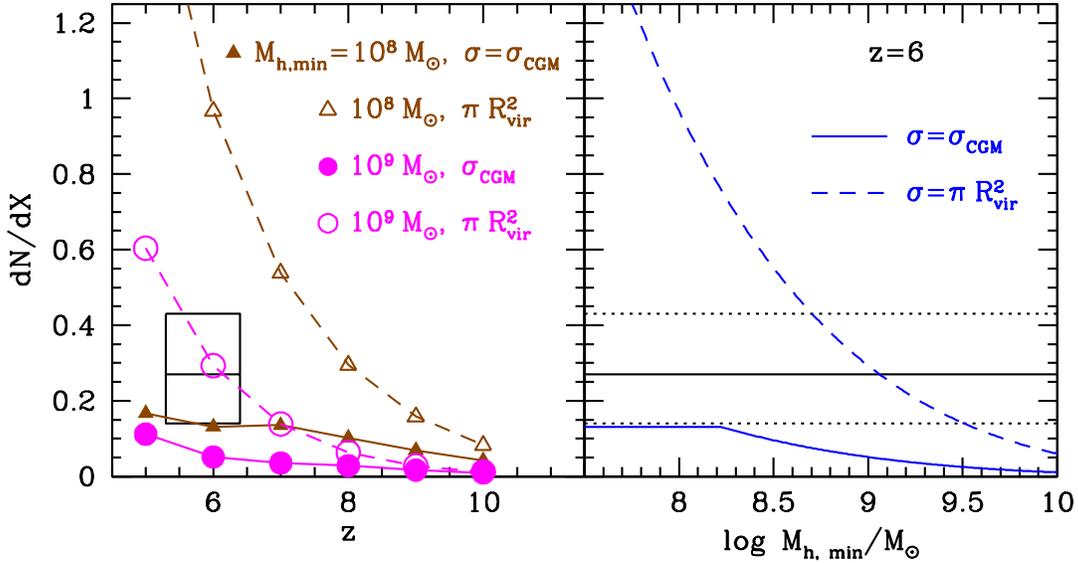}}
}
\vspace{-3in}
\caption{
(left) The number density of absorbers with columns greater
than $10^{14}$cm$^{-2}$ as a function of redshift.  Filled brown 
triangles represent our predictions, which are marginally consistent 
with the observed 95\% confidence interval of~\citet{bec11} (box).  
Filled magenta circles use the same cross sections, but integrating 
only down to $10^9\msun$.  Open triangles and circles indicate the 
$\rvir$ model, integrated down to $10^8$ and $10^9\msun$, 
respectively. (right) The dependence of the predicted abundance 
at $z=6$ on a hypothetical low-mass cutoff in our simulations (solid ) 
and in the $\rvir$ model (dashed).   Horizontal lines indicate the 
observed 95\% confidence interval.
}
\label{fig:dNdX}
\end{figure*}

By integrating Equation~\ref{eqn:dldM} over halo mass, we may compute 
the 
predicted number of absorbers per absorption path length.  In the 
left panel of Figure~\ref{fig:dNdX}, we compare the predicted and 
observed abundances as a function of redshift.  The observations 
are from~\citet{bec11}, who identified nine OI absorbers along a 
total path length $\Delta X = 39.5$
between $z=5.3$ and $z=6.4$.  They estimate that, for systems with
columns in excess of $10^{14}$cm$^{-2}$, their observations are 
80--85\% complete.  Correcting for an assumed 85\% completeness, 
we estimate an observed abundance of $0.27^{+0.16}_{-0.13}$ absorbers 
per path length with columns greater than $10^{14}$cm$^{-2}$, 
where the confidence intervals are 95\% and account only for 
Poisson uncertainty.

The predicted absorber abundance (solid brown curve with filled triangles) 
is in marginal agreement with the
observational $2\sigma$ confidence range.  This level of consistency 
is remarkable given that the simulation has been calibrated 
using observations of high-ionization metal absorbers at lower 
redshifts~\citep{opp06} and tracers of hydrogen reionization 
(Section~\ref{ssec:sims}).  The implication is that low-ionization 
metal absorbers are complementary probes of the same physical processes.  

In detail, the predicted abundance of OI absorbers
lies just below the observed 2$\sigma$ confidence intervals 
at $z\approx6$ (open box).  At this point, it is interesting to recall 
that our simulations also slightly overpredict the amplitude of the 
EUVB at $z=6$ (Section~\ref{ssec:reionhist}).  These inconsistencies are 
probably telling the same story: At $z=6$, the simulated EUVB is
too strong, yielding an optical depth to Lyman-$\alpha$ absorption that
is too low.  For the same reason, the simulated ionization fronts 
penetrate too far into halos, yielding geometric cross sections for
low-ionization absorption that are too small, hence the predicted 
abundance of neutral metals is also low.

An additional source of uncertainty is the assumed metal yields:
For reasonable choices of initial mass function and Type II supernova
yields, the total oxygen yield can vary by a factor of 2--3.  
Doubling the assumed oxygen yield would not violate constraints on the 
$\NOI/\NHI$ ratio (Figure~\ref{fig:ulasj1120}), but it would boost 
the predicted cross sections into improved agreement with observations.

The abundance is predicted to evolve quite slowly owing to 
cancellation between the growing abundance of halos and their declining 
geometric cross sections.  If this evolution continues to lower redshifts,
then it readily explains the slow evolution in the observed abundance of 
low-ionization absorbers between $z=6$ and $z=3$~\citep{bec11}.

Our simulations could underestimate the rate at which gas is expelled from 
photosensitive halos owing to incorrect outflow scalings or an improper 
treatment of the radiation field on small scales (see below), hence we 
recompute the abundance omitting halos less massive 
than $10^9\msun$ and show the result using filled magenta circles.  This 
toy model lies well below observations, confirming that further suppression 
of star formation in photosensitive halos cannot be accommodated by existing data
unless the EUVB is significantly weaker.  As there is nothing special about 
$10^9\msun$, we show in the right panel the predicted abundance at $z=6$ as 
a function of the cutoff mass (solid blue curve); this figure confirms that 
the true cutoff mass cannot be much higher than $10^{8.5}\msun$ at 
$z=6$.  Open triangles in the left panel show the $\rvir$ model using halos 
more massive than $10^8\msun$.  This comparison shows that absorption in 
$10^8\msun$ halos cannot extend out to the virial radius at $z=6$, and our 
simulations provide a self-consistent model for how this occurs.
Open circles show the $\rvir$ model assuming that halos below 
$10^9\msun$ do not contribute at all.  This model is consistent 
with observations at $z=6$, suggesting that $10^9\msun$ halos host 
OI absorbers.  Within our simulations, this is in fact the dominant
mass scale (Figure~\ref{fig:MfromN}).  It falls below the simulated 
abundance at $z>7$, indicating the growing role that photosensitive halos 
may play at higher redshifts.

In summary, weighting the dark matter halo mass function by analytic fits 
to the predicted median trend of cross section versus halo mass confirms that
OI absorption is dominated by the lowest-mass halos that sustain star 
formation at any redshift.  The predicted absorber abundance evolves slowly 
owing to strong cancellation between the growing halo abundance and the 
declining cross section at a given mass, and it is in marginal agreement 
with observations at $z=6$.  In detail, however, the predicted abundance
is slightly low, consistent with the fact that the amplitude of the predicted
EUVB is slightly too high.

\subsection{Overlap with Neutral Hydrogen Absorbers}\label{ssec:HI}
\begin{figure}
\centerline{
\setlength{\epsfxsize}{0.5\textwidth}
\centerline{\epsfbox{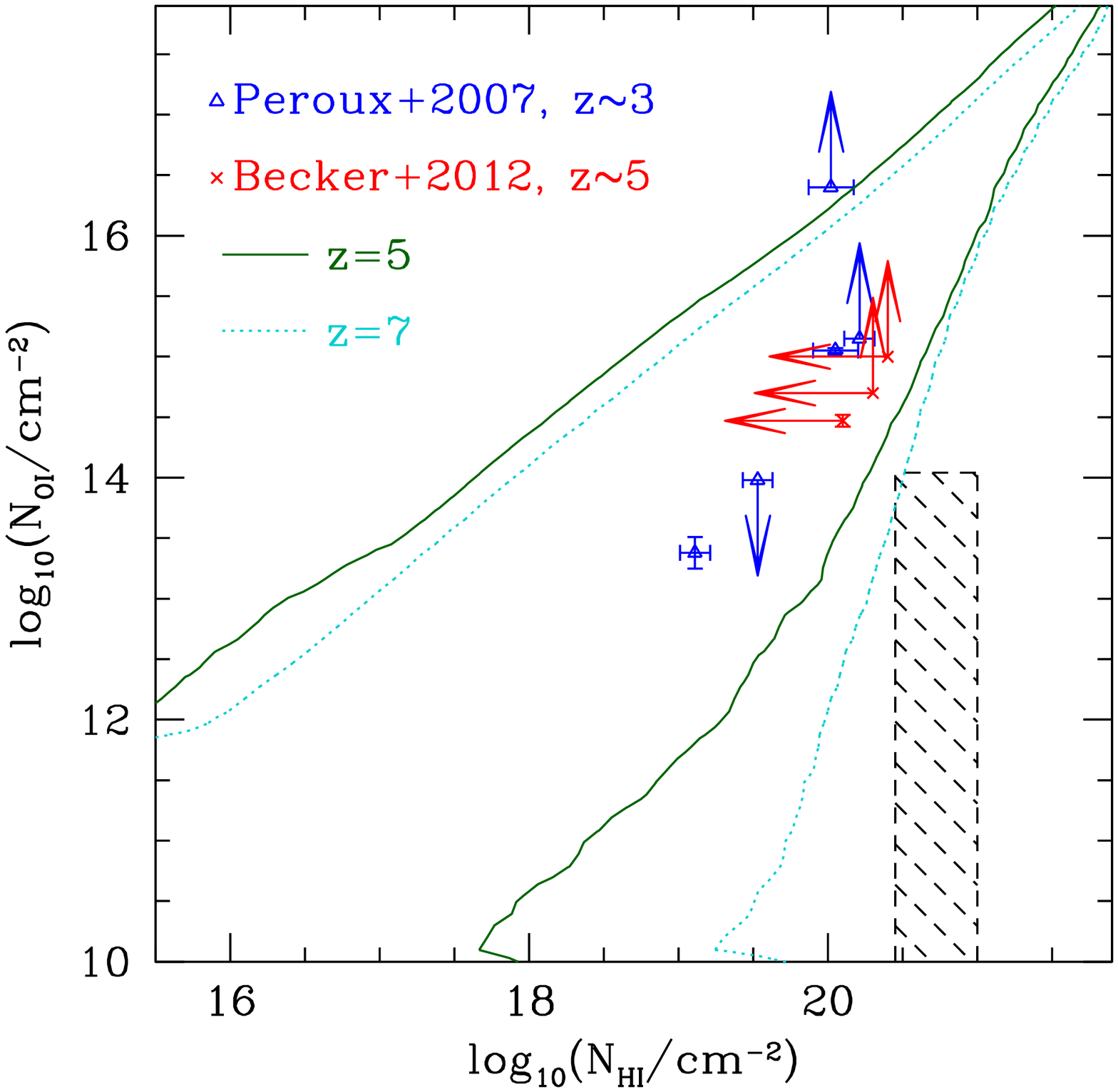}}
}
\caption{The dependence of neutral oxygen column density on the neutral
hydrogen column density along lines of sight that pass through halos 
at $z=7$ (dotted turquoise) and $z=5$ (solid green).  Contours enclose 99\%
of all sightlines.  Metal absorbers with column densities
$\log_{10}(\NOI)\geq14$ correspond to neutral hydrogen columns of 
$\log_{10}(\NHI)=17-21.5$.  Blue triangles are HI-selected absorbers
at $z\sim3$~\citep{per07} while red crosses correspond to 
$z\sim5$~\citet{bec12}.  The overlap with the predicted abundance ratios 
indicates agreement between predicted and observed metallicities.
The shaded region shows the metallicity constraints
on the foreground absorber in front of ULASJ1120+0641 under the assumption
that it is gravitationally bound~\citep{sim12}.  Halo gas is already too 
enriched at $z=7$ to satisfy these constraints, suggesting that the absorber 
lies in the diffuse IGM.
}
\label{fig:ulasj1120}
\end{figure}
The conclusion that low-ionization metal absorbers correspond to bound gas
raises the question of how they relate to more familiar absorption-selected 
populations at lower redshifts.  In particular,~\citet{bec11,bec12} suggested 
that low-ionization metal absorbers could be analogous to DLAs and sub-DLAs.
In order to consider 
this possibility, we project halos onto a grid with cells 100 physical pc wide 
(which is roughly the gravitational softening length) and pass sightlines 
through each pixel that falls within one virial radius in order to compute
the neutral hydrogen and oxygen columns.  We grid each halo in the x, y, and z 
directions independently in order to account for departures from spherical 
symmetry.  We show how the $\NOI$ and $\NHI$ columns compare in 
Figure~\ref{fig:ulasj1120}.  Contours enclose 99\% of all sightlines.  

At $z=7$, observable OI absorption ($\NOI>10^{14}$cm$^{-2}$) can arise in 
systems with neutral hydrogen columns of 
$10^{18}$--$10^{21}$cm$^{-2}$.  By $z=5$, ongoing metal enrichment boosts
the typical $\NOI$ as a function of $\NHI$ so that OI absorbers can be found
in weaker systems.
Additionally, the population of absorbers with relatively 
low metal columns (that is, the ``tail" to low $\NOI/\NHI$) contracts, 
reflecting rapid enrichment of moderately overdense gas owing to 
low-mass systems.

\citet{bec12} have measured the OI and HI column for three low-ionization
systems at $z\sim5$; we include results from their Table 2 using red 
crosses.  The overlap with the predicted abundance ratios at $z=5$ is
excellent, indicating that the simulated metallicities are reasonable.

In order to emphasize the identification between OI-selected systems in our
simulations and HI-selected systems from observations at lower redshifts, we
also include OI constraints on DLAs and sub-DLAs at $z\sim3$ 
from~\citet{per07} (blue triangles).
These measurements span the predicted range at $z=5$, suggesting that the
evolution in the metal mass fraction at given HI column is comparable
to the scatter at fixed redshift.  Unfortunately, we cannot compare 
predictions and observations of HI-selected systems (note that 
the~\citealt{bec12} systems are selected as metal absorbers) directly because 
we have not evolved our simulations past $z=5$.   Nevertheless, 
Figure~\ref{fig:ulasj1120} clearly supports the suggestion by~\citet{bec12} 
that systems selected in low-ionization metal transitions are physically 
analogous to HI-selected systems.

\begin{figure}
\centerline{
\setlength{\epsfxsize}{0.5\textwidth}
\centerline{\epsfbox{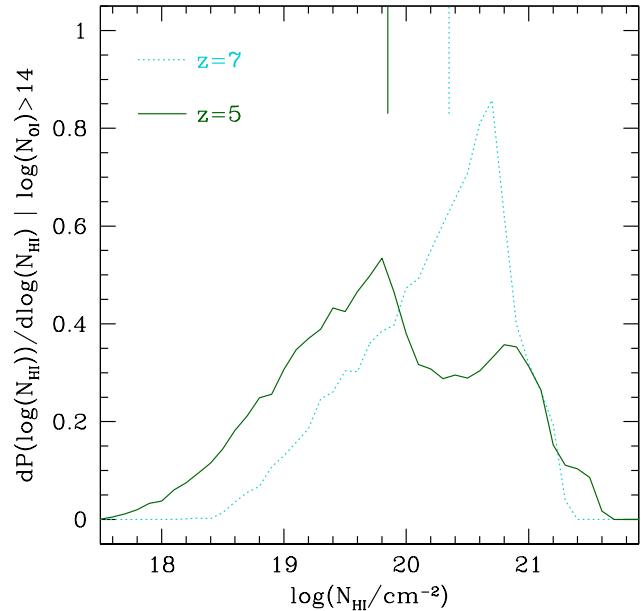}}
}
\caption{The distribution of neutral hydrogen columns for low-ionization
metal absorbers with column densities in excess of $10^{14}$cm$^{-2}$
at $z=7$ (dotted turquoise) and $z=5$ (solid green).  Vertical segments indicate
the median $\NHI$.  Low-ionization metal absorbers receive significant 
contributions from DLAs and sub-DLAs, with a tail extending to LLS columns
by $z=5$.
}
\label{fig:pNHIgNOI}
\end{figure}
By selecting only systems with $\NOI>10^{14}$cm$^{-2}$, we may ask directly
what the predicted distribution of neutral hydrogen columns of low-ionization
metal absorbers is.  We show these probability distribution functions at 
$z=7$ (from our fiducial simulation) and $z=5$ (from the r6n256wWwRT16d
simulation, which has the same physical treatments but subtends 
$(6/9)^3\approx0.3$ times the cosmological volume) in Figure~\ref{fig:pNHIgNOI}.  
The vertical segment shows the median $\NHI$.  At $z=7$, 
roughly half of OI absorbers are DLAs while half are sub-DLAs.  By $z=5$, 
the fractional contribution from DLAs with $\NHI\sim10^{21}$cm$^{-2}$ 
remains unchanged.  Meanwhile, ongoing enrichment boosts the neutral metal 
columns of systems $\NHI<10^{20}$cm$^{-2}$.  This suppresses the median 
$\NHI$ of metal-selected samples into the sub-DLA range.  Unfortunately, the 
doubly-peaked probability distribution function at $z=5$ indicates that 
our predictions suffer from volume limitations.  Broadly, however,
the conclusion is clear: Systems that are selected to have 
$\NOI>10^{14}$cm$^{-2}$ are drawn with roughly equal probability from DLAs 
and sub-DLAs, with a slight evolution to lower hydrogen columns at lower 
redshifts owing to ongoing enrichment.  Note that the distribution at $z=5$
as well its inverse, $\NOI(\NHI)$ are predictions that may be tested 
directly.


Figure~\ref{fig:ulasj1120} also suggests that the overall number of detected
systems increases for lower threshold column densities.  This is not 
surprising;~\citet{opp09} predicted that the column density distribution 
$d^2N/dXdN$ for low-ionization metals varies with $N$ as $N^\alpha$, where 
$\alpha$ falls between -1 and -2.  While a detailed discussion of the
predicted OI column density distribution is beyond the scope of the current
work, we have recomputed the number of absorbers per path length
$dN/dX$ for different threshold column densities using only the halos that 
occur in our fiducial simulation at $z=6$ (that is, there is no correction
for more massive halos that are undersampled owing to volume limitations).  
Taking the ratio with respect to the $dN/dX$ for systems with $\NOI > 10^{14}$ 
cm$^{-2}$, we find relative abundances of 1.3, 1.1, and 0.8 for systems 
with $\NOI$/cm$^{-2}$ $ > 10^{13}$, $5\times10^{13}$, and $5\times10^{14}$, 
respectively.  Hence we expect a modest increase in the overall number of 
systems identified as survey sensitivity limits improve.

\subsection{Foreground Absorption in ULAS J1120+0641}\label{ssec:ulasj1120}

The $z=7.085$ quasar ULASJ1120+0641, reported by~\citet{mor11}, shows 
strong foreground absorption that
could owe either to gas that lies within the quasar's host galaxy, to an 
unassociated gravitationally-bound system analogous to DLAs that lies in the
foreground, or to a neutral patch of the IGM that happens to lie along the 
line of sight.
Recently,~\citet{sim12} used a high-resolution infrared spectrum to search 
for metal absorption at the position of the foreground absorber.
The detection of metal absorption would give strong support to the view 
that it corresponds to a discrete system rather than to the diffuse IGM.  
They estimated a foreground neutral hydrogen column of 
$\log_{10}(\NHI/\mathrm{cm}^{-2})=$20.45--21.0.  They did not 
detect any metal absorption at the redshift of the absorber (with the 
exception of a 2.2$\sigma$ detection of neutral oxygen).  Modeling their
upper limits under the assumption that the absorber was discrete, they 
found that the OI column density was constrained to be less than 
$14.04$ cm$^{-2}$ at $2\sigma$.  

We may use our simulated sightlines to ask whether these measurements are 
consistent with arising in halo gas.  To this end, we compare in 
Figure~\ref{fig:ulasj1120} the predicted distribution of OI and HI columns
at $z=7$ (cyan contours) with the observationally-allowed combinations 
(shaded region).  Given that 1\% of simulated sightlines lie outside of
the cyan region, it is clear at a glance that there is negligible overlap 
between the predicted column density ratios of halo gas and the~\citet{sim12}
constraints.  For completeness, however, we may estimate the predicted
probability that the observed absorption does arise in halo gas as follows:
The redshift of the absorber is constrained to be $7.041\pm0.003$, 
corresponding to a 3-$\sigma$ observed path length of 0.79 Mpc.
Neglecting shadowing and including sightlines in the x, y, and z directions, 
our simulations effectively model a path length of $3.06\times10^8$ Mpc and
identify 34,090 absorbers that satisfy the~\citet{sim12} constraints.  Hence
the model predicts that the probability of encountering a system satisfying
the~\citet{sim12} constraints over the observed path length is $\sim10^{-4}$.
We conclude that the foreground absorber is not consistent with arising in 
halos because bound gas at the inferred HI column is already too enriched 
at $z=7$.

If the foreground absorption does not arise in a halo, then it must arise
in the diffuse IGM.  In this case,~\citet{sim12} find that the metal mass 
fraction must be less than $10^{-3}$ in solar units.  In our simulations, 
the mean metallicity at $z=7$ falls below this limit for all gas that is 
less dense than 0.3 times the mean density.  In other words, a typical 
underdense region readily satisfies the observational constraints.

We may also consider under what conditions the foreground absorption could 
originate in bound gas without matching the OI column densities expected 
from our
simulations.  The class of simulations presented here yields reasonable 
agreement with a wide range of observational constraints including the 
galaxy mass-metallicity relation at $z=2$~\citep{fin08}, the abundance of 
CIV absorbers at $z=6$~\citep{opp09}, the rest-frame ultraviolet luminosity 
function of galaxies~\citep{dav06,opp09,fin11b}, and the history of 
reionization (Section~\ref{ssec:reionhist} and~\citealt{fin12}).  Thus,
we expect the OI column densities to be a robust prediction.  Nonetheless, 
the foreground absorption could still originate in bound gas if the true
star formation rate density per unit gas density $\rhodotstar/\rhogas$ is 
lower than assumed in our models.  In this case, gas could remain 
unenriched until it reaches much 
higher columns than $\NHI=20.5$ cm$^{-2}$.  A dependence 
of $\rhodotstar/\rhogas$ on metallicity is in fact expected
theoretically~\citep{gne10}, and implementations of this idea within
numerical simulations generically predict suppressed star formation
rates in low-mass halos~\citep{kuh12b,tho13}.  We will return
to these models in Section~\ref{sec:discuss}; for now, we note that this 
possibility motivates future work comparing the dependence of metallicity 
on density in simulations that assume different star formation and 
feedback models.

In summary, the stringent limits on the metal abundance of the absorbing
gas are not satisfied by overdense gas in our simulations.  Meanwhile, 
they are readily satisfied by underdense gas at $z\geq7$.  Given the 
level of realism implied by the range of observational constraints
that our simulations are known to satisfy, these comparisons argue that 
the intervening gas lies in the diffuse IGM rather than in a discrete 
absorber.  This implies a volume-averaged neutral hydrogen fraction of
$\sim10\%$ at $z=7$~\citep{bol11,sch13}.

\section{Discussion}\label{sec:discuss}

\subsection{Implications}
The agreement between the predicted and observed abundances of 
OI absorbers in Figure~\ref{fig:dNdX} is consistent with a scenario 
in which star formation persists 
at scales a factor of 10--100 lower in mass than is currently probed by 
observations of galaxies in emission~\citep{mun11,ouc10}.  If true, then
the abundance of OI absorption systems places a strong constraint on 
models in which star formation in low-mass halos is suppressed owing to, 
for example, inefficient formation of molecular clouds at low 
metallicities~\citep{kru12,kuh12b,chr12,tho13} or a mass 
threshold below which gas accretion ceases~\citep{bou10}.  

At the same time, the abundance of neutral oxygen must be suppressed 
in halos roughly a factor of 10 more massive than the hydrogen 
cooling limit by $z=6$ or else the abundance of OI absorbers 
would be substantially overproduced (brown dashed curve with open 
triangles).  Given that such systems readily form stars~\citep{fin11b} 
and enrich their CGM prior to the onset of reionization 
(Figure~\ref{fig:rpmassz10}), this indicates that their CGM 
must be substantially ionized at $z=6$.

As an alternative to this picture, it is possible that star formation
is inefficient in halos much more massive than $10^9\msun$ as long
as the cross-section for more massive halos to appear as OI absorbers
significantly exceeds their virial radius.  For example,~\citet{kuh12b}
have shown that a model in which stars form only out of molecular 
gas suppresses star formation in halos less massive than $10^{10}\msun$ at 
$z\geq4$.  If true, this model would imply that OI absorbers are associated
with larger halos than predicted by our simulations.  They could not be
arbitrarily large, however: The velocity widths reported by~\citet{bec11} 
are narrower than would be expected for gas associated with $10^{11}\msun$ 
halos, with five out of seven systems exhibiting velocity widths of less 
than $100\kms$.  

We disfavor this option for two reasons.  First, the model discussed
by~\citet{kuh12b} may be too efficient at suppressing star formation in
low-mass galaxies.  For example, it underproduces the abundance of faint 
UV-selected galaxies at $z=4$ (see the KMT07 curve in Figure 18).  
Similarly, recent observations suggest that the UV luminosity function 
rises to -13 at $z=2.4$~\citep{ala13}, in conflict with predictions 
from metallicity-dependent cooling models~\citep{kuh13}.  The results 
of~\citet{ala13} are based on only a few objects, but if verified in 
future work then they indicate that star formation continues in halos 
less massive than $10^{10}\msun$, in agreement with our predictions.  

These discrepancies may be removed by increasing the mass 
resolution~\citep[Figure 16 of][]{kuh12b} or changing the numerical 
implementation~\citep[for example,][]{jaa13,tho13}, although in all
of these models the predicted UV luminosity function turns over at
a luminosity is too bright to match the~\citet{ala13} results.
Increasing the efficiency of star formation in low-mass systems 
sufficiently to reproduce the abundance of faint UV-selected galaxies
would decrease the halo mass below which star formation is suppressed,
bringing these models back into agreement with ours.  

More broadly, the assumption that reionization was driven 
by galaxies already leads naturally to the conclusion that star 
formation must continue to scales at least $100\times$ fainter than 
current limits.  This idea is further supported both by considerations
regarding the number of ionization photons that can be provided by 
observed galaxies~\citep{rob13,cal13} as well as the fraction of gamma-ray
bursts with no optical counterpart~\citep{tre12}.  The view that OI 
absorption directly probes this population is more natural.  

The second difficulty with models attributing OI absorbers to halos more
massive than $10^{10}\msun$ is that it is difficult to understand 
how an optically-thick, enriched CGM would extend with large cross-section
to such large distances around massive halos given that they are 
expected to live in regions where the EUVB is more intense.  
Even at $z=10$, halos are not completely optically thick in our 
simulations (Figure~\ref{fig:sigvmass}), hence it is not likely that 
a significant absorption column exists well outside the virial radius.

The idea that OI absorbers originate in low-mass halos raises the
possibility of using absorbers, Lyman break galaxies, and Lyman-$\alpha$ 
emitters jointly to constrain how star formation and feedback scale 
with halo mass across a much wider range of halo 
masses than can be probed by any population alone.  Such an inquiry 
would require an improved understanding of the connection 
between a halo's star formation rate and its cross section for 
observability in absorption, a daunting undertaking both for theory
and for observation~\citep{fyn08,kro13}.  However, the 
reward would be a powerful probe of star formation and feedback 
across many decades of dynamic range.

Our simulations predict that the overall abundance of OI absorbers
increases slowly in time, particularly below $z=6$.  
It should be possible to test this prediction with existing 
observations, although existing catalogs of absorbers tend to be 
pre-selected as DLAs rather than as OI absorbers; the overlap 
between these two populations would require improved understanding 
in order to correct for selection biases.

\subsection{Limitations}
Our simulations suffer from several limitations associated with 
resolution and numerical methodology.
First, they resolve halos at the hydrogen cooling limit with
$\sim100$ particles.  While this is sufficient for a converged mass 
density profile~\citep{tre10}, it is not clear that this criterion
also leads to a converged absorption cross-section.  To explore this,
we compared the absorption cross-section for two simulations with
different mass resolutions.  These simulations, the r3wWwRT32 and 
r6wWwRT32 simulations from~\citet{fin11b}, use $2\times256^3$ 
particles to model 3 and $6\hmpc$ volumes, respectively.  In
contrast to our more recent simulations, they adopt a constant 
$\fesc=0.5$ and do not include a subgrid self-shielding prescription.  
To simulate self-shielding in post-processing, we therefore assume 
that all gas with baryon overdensity greater than 320 is fully 
neutral while less dense gas is fully ionized; this approximates
the behavior of our more recent simulations at $z=$6--7.  For 
halos more massive than $10^9\msun$, the cross sections in the 
high-resolution calculation are roughly 70\% as large as at our 
fiducial resolution at $z=6$ and $z=7$.  The difference owes 
to the explicit dependence of outflow properties on halo mass 
coupled with the fact that star formation begins sooner at higher 
mass resolution.  Hence we estimate that mass resolution limitations 
affect the predicted abundance of absorbers at the $\sim50\%$ level.

Second, our simulations do not treat the interaction between outflowing 
gas and the CGM correctly because outflows consist of isolated gas particles 
that are expelled at roughly the escape velocity from star-forming regions.  
Given that smoothed-particle hydrodynamics defines a particle's thermal 
properties by smoothing over the properties of neighboring particles, this 
simplified treatment precludes the formation of multiphase outflows in which 
cold, optically thick cores are entrained in a hot, optically-thin medium.  
The lack of a cold component embedded in the outflows could in turn lead us 
to underestimate the geometric cross section to absorption in low-ionization 
transitions.  For similar 
reasons, our simulations probably do not treat the mixing that occurs
between different phases in the CGM correctly.  This effect may be crucial
in reconciling models with the observed velocity width distribution of
DLAs~\citep{tes09}.

Third, our simulations are known to treat shocks and 
hydrodynamic instabilities inaccurately, leading to unphysical behavior 
at fluid boundaries.~~\citet{bir13} have compared the $f(\NHI)$ predictions
in simulations using smoothed particle hydrodynamics versus a new 
moving-mesh formalism, {\sc arepo}, that alleviates many of these issues.  
{\sc arepo} predicts that absorbers with $\NHI=10^{19}$--$10^{20}$ are
more abundant than in {\sc gadget} whereas absorbers with 
$\NHI=10^{20}$--$10^{21}$ are less abundant.  Coincidentally,
our simulations predict that OI absorbers fall with roughly equal 
probability into these two ranges (Figure~\ref{fig:pNHIgNOI}), hence
the overall impact on the predicted OI absorber population is difficult
to predict.  Moreover, their calculations did not include a treatment 
for galactic outflows, which can significantly impact absorption 
statistics~\citep{nag07,tes09}.  A more complete appraisal of
the impact of hydrodynamic instabilities will therefore require 
further work.

Fourth, our model neglects ionizations that occur within halos owing to
nearby stars.  This is because it attenuates the radiation field on scales 
smaller than the radiation transfer grid using a subgrid self-shielding 
approach.  Implicitly, our simulations assume that a fraction $1-\fesc$ of 
ionizing photons are absorbed by molecular clouds that behave as photon sinks 
while the other $\fesc$ escape through optically-thin holes directly into the 
IGM, where the mean free path is large enough to be resolved by our radiation
transport solver.  This yields a purely outside-in reionization 
topology~\citep{mir00} in which gas that is more dense than an evolving
threshold density is completely neutral.  Analytic estimates suggest that 
local ionizations could suppress the abundance of neutral hydrogen 
absorbers with columns greater than 
$10^{17}$cm$^{-2}$~\citep{mir05,sch06} at $z=3$, which in our model 
includes all observable OI absorbers (Figure~\ref{fig:pNHIgNOI}).  
Indeed, the local field could 
be even more important at $z\geq6$, when the EUVB is much weaker.  

While a detailed calculation of the local field is beyond the scope of our 
current work, it is useful to consider the implications of recent studies.
~~\citet{nag04,nag07} used a subgrid prescription for the 
multiphase ISM~\citep{spr03} to model the ISM neutral fraction and found 
that the abundance of neutral hydrogen absorbers $f(\NHI)$ with columns 
$\NHI<10^{21}$cm$^{-2}$ at $z=3$ was underproduced; this suggested that dense
gas was overionized.~~\citet{tes09} reproduced their result using a similar 
prescription and found that, by assuming that all gas more dense than a
threshold of 0.01cm$^{-3}$ was neutral, they could increase 
the predicted DLA abundance by 0.2 dex.~~\citet{nag10} confirmed that invoking a 
slightly lower density threshold ($6\times10^{-3}$cm$^{-3}$) yielded excellent 
agreement with observations across a wide range of 
$\NHI$.~~\citet{pon08},~\citet{mcq11}, and~\citet{yaj12}, used radiation 
transport calculations to calculate the 
threshold density and found excellent agreement with the observed $f(\NHI)$.  
These works indicate that the local field must be modest within the halos 
that dominate $f(\NHI)$ at $z=3$.

At the factor-of-two level, however, it cannot be ignored.~~\citet{yaj12} 
found that the local radiation field reduces the geometric absorption cross 
section for $10^9\msun$ halos by $\approx50\%$ at $z=3$.~~\citet{rah13} found 
that the local field suppresses the abundance of systems with 
$\NHI=10^{19}$--$10^{21}$cm$^{-2}$ by a factor of 3 at $z=5$.  Importantly, 
they also noted that the role of the local field is quite sensitive to the 
uncertain relative spatial distribution of sources and sinks within the ISM.  
Finally,~\citet{fum11} found that the local background reduces the cross 
section by no more than $50\%$ for absorbers with columns 
$\NHI=10^{18}$--$10^{21}$cm$^{-2}$ (their Figure A1).  These studies, many
of which incorporate much higher resolution than ours, suggest that the local
field can suppress $f(\NHI)$ by a factor of 2--3.  Given the tight coupling
between the hydrogen and oxygen neutral fractions, we conclude that it could
likewise suppress the OI cross sections by a factor of 2--3.  This crude 
estimate neglects the role of radial metallicity gradients, but it indicates 
that uncertainties associated with the local field are not large compared to 
uncertainties associated with the limited observational sample size.  On
the other hand, the local field's significance may be comparable to the 
amount by which cross sections shrink owing to the strengthening EUVB 
(Figure~\ref{fig:sigvMzconst}).

Another consequence of the local ionizing background could be to modulate
the dominant mass scale of OI absorbers' host halos.  In particular, if
$\fesc$ increases to low halo masses~\citep{alv12,yaj11}, then it could suppress 
the absorption
cross-section in halos less massive than $10^9\msun$.  In order not to
compromise the good agreement between the predicted and observed abundance
of absorbers (Figure~\ref{fig:dNdX}), a small decrease in the cross-section 
of low-mass halos would have to be compensated by a large increase in 
halos with masses in the range $10^9$--$10^{11}\msun$ (more massive host 
halos would be difficult to reconcile with the velocity widths reported 
by~\citealt{bec11}), but this does not seem impossible.   We conclude that,
while our simulations represent a plausible model for low-ionization 
absorbers, the local ionizing background remains a source of uncertainty
regarding the nature of their host population.

Finally, our model for galactic outflows may not incorporate the correct 
scaling between outflow mass loading factors, velocities, and host galaxy
properties.  This is important because the properties of DLAs are
sensitive to outflows~\citep{nag07,tes09}.  It has recently been shown 
that an alternative model in which low-mass galaxies eject more mass per 
unit stellar mass formed than in the current model produces improved 
agreement with the observed mass function of neutral hydrogen~\citep{dav13}.  
This model may predict lower metallicity in dense gas, suppressing the 
abundance of low-ionization metal absorbers at high column densities.
while enhancing the abundance of high-ionization absorbers.

In summary, mass resolution limitations may affect the predicted
abundance of OI absorbers at the $\sim50\%$ level; the local ionizing 
background could affect results at the $\sim$factor of 3 level; and 
the impact of uncertainties related to inaccuracies in our hydrodynamic 
solver, our treatment for galactic outflows, and $\fesc$ is difficult 
to ascertain.  None of these considerations challenges the basic 
prediction that OI absorbers are associated with gravitationally 
bound gas in systems that are analogous to DLAs and sub-DLAs at 
lower redshifts.  Further work is required, however, in order to 
improve our understanding of the likely mass scale of their host 
halos.

\section{Summary}\label{sec:sum}
We have used a cosmological radiation hydrodynamic simulation to study the
nature of OI absorption in the reionization epoch.  The diffuse IGM is not
sufficiently enriched by $z=10$ for OI to trace its ionization 
state directly.  Instead, OI is tightly associated with dense gas that lies
within dark matter halos.  In the absence of an EUVB, all halos more 
massive than the hydrogen cooling limit possess significant reservoirs of
neutral oxygen out to a substantial fraction of the virial radius; in this 
case OI observations are dominated by halos near the hydrogen cooling 
limit.  An EUVB ionizes and evaporates gas out of halos less 
massive than $\sim10^{9}\msun$; such halos then become unobservable 
at their virial radius with the result that the characteristic host 
halo mass of low-ionization absorbers increases.  This may cause the 
characteristic velocity widths to increase to lower redshifts.  Even 
so, however, the dominant host halos of OI absorbers are not more massive 
than $10^9\msun$ at $z=6$.  Hence OI absorbers trace the signatures of 
star formation in halos a factor of 10--100 less massive than the halos
that host emission-selected samples such as Lyman break galaxies and
Lyman-$\alpha$ emitters.  Our simulations yield marginal 
agreement with the observed OI absorber abundance at $z\sim6$.  In
detail, the predicted abundance is slightly low, consistent with the
fact that the predicted EUVB is slightly too strong compared to 
constraints from the Lyman-$\alpha$ forest.

We additionally compare our density profiles to the upper limits on the 
OI column density of the absorber in the foreground of the $z=7.085$ quasar 
ULASJ1120+0641 and find that the limits cannot be satisfied by gas within halos 
because gas at the observed HI column density 
is already too enriched by $z=7$.  By contrast, gas at less than one third 
the mean density has a low enough metallicity to satisfy the constraints at 
$z\geq7$.  This supports the view that the absorption occurs in the diffuse 
IGM rather than in a discrete system and argues for a $\sim10\%$ 
volume-averaged neutral hydrogen fraction.

\section*{Acknowledgements}
We thank R.\ Somerville, M.\ Peeples, and M.\ Prescott for helpful 
conversations.  We also thank the anonymous referee for many 
thoughtful suggestions that inspired improvements to the draft.
As always, we are indebted to V.\ Springel for making {\sc Gadget-2}
available to the public.  Our simulations were run on the University 
of Arizona's Xeon cluster and on TACC's Ranger supercomputer.
Support for this work was provided by the NASA Astrophysics Theory 
Program through grant NNG06GH98G, as well as through grant number 
HST-AR-10647 from the SPACE TELESCOPE SCIENCE INSTITUTE, which is 
operated by AURA, Inc. under NASA contract NAS5-26555.  Support for 
this work, part of the Spitzer Space Telescope Theoretical Research 
Program, was also provided by NASA through a contract issued by the 
Jet Propulsion Laboratory, California Institute of Technology under 
a contract with NASA.  KF gratefully acknowledges support from NASA 
through Hubble Fellowship grant HF-51254.01 awarded by the Space 
Telescope Science Institute, which is operated by the Association 
of Universities for Research in Astronomy, Inc., for NASA, under 
contract NAS 5-26555.  KF thanks the Danish National
Research Foundation for funding the Dark Cosmology Centre.
SPO acknowledges support from NASA grant NNX12AG73G.


\onecolumn

\begin{deluxetable}{l|cccc}
\tabletypesize{\footnotesize}
\tablecolumns{5}
\tablewidth{0pt}
\tablecaption{Our simulations.  The fiducial simulation is
indicated in bold.\label{table:sims}}
\tablehead{
\colhead{name} &
\colhead{$L$\tablenotemark{a}} &
\colhead{RT grid} &
\colhead{outflows?} &
\colhead{self-shielding?}
}
\startdata
r6n256wWwRT16d    & 6       & $16^3$ &	yes    	& yes 	\\
r6n256nWwRT16d    & 6	& $16^3$ &	no 	& yes 	\\
{\bf r9n384wWwRT48d} & {\bf 9}      & $\mathbf{48^3}$ &	{\bf yes}   	& {\bf yes} \\ 
r6n256wWwRT	      & 6	& $16^3$ &	yes 	& no 

\enddata
\tablenotetext{a}{in comoving $\hmpc$}
\end{deluxetable}

\begin{deluxetable}{l|ccccc}
\tabletypesize{\footnotesize}
\tablecolumns{6}
\tablewidth{0pt}
\tablecaption{Cross-section versus halo mass and redshift.\label{table:fits}}
\tablehead{
\colhead{redshift} &
\colhead{$a_l$} &
\colhead{$b_l$} &
\colhead{$a_h$} &
\colhead{$b_h$} &
\colhead{$\log_{10}(M_{h,c}/\msun)$}
}
\startdata
5  &  -535.981  &  63.5  &  -6.8  &  0.875  &  8.45  \\
6  &  -554.636  &  67.5  &  -5  &  0.675  &  8.225  \\
7  &  -483.55  &  60.5  &  -4.95  &  0.675  &  8  \\
8  &  -422.295  &  53  &  -6  &  0.8  &  7.975  \\
9  &  -238.141  &  30  &  -6.2  &  0.825  &  7.95  \\
10  &  -236.74  &  30  &  -6.85  &  0.9  &  7.9  \\
\enddata
\end{deluxetable}

\end{document}